\documentclass[pdflatex]{sn-jnl}

\usepackage[T1]{fontenc}
\usepackage{upgreek}

\usepackage{graphicx}%
\usepackage{multirow}%
\usepackage{amsmath,amssymb,amsfonts}%
\usepackage{amsthm}%
\usepackage{mathrsfs}%
\usepackage[title]{appendix}%
\usepackage{xcolor}%
\usepackage{textcomp}%
\usepackage{manyfoot}%
\usepackage{booktabs}%
\usepackage{algorithm}%
\usepackage{algorithmicx}%
\usepackage{algpseudocode}%
\usepackage{listings}%

\newcommand{\reffig}[1]{Fig. \ref{#1}}

\newcommand{\refequ}[1]{Eq. (\ref{#1})}

\newcommand{\KB}[0]{\rm k_{B}}

\begin{document}

\title[Entropy scaling for diffusion coefficients in fluid mixtures]{Entropy scaling for diffusion coefficients in fluid mixtures}

\author[1]{\fnm{Sebastian} \sur{Schmitt}}\email{sebastian.schmitt@rptu.de}

\author[1]{\fnm{Hans} \sur{Hasse}}\email{hans.hasse@rptu.de}

\author*[1]{\fnm{Simon} \sur{Stephan}}\email{simon.stephan@rptu.de}

\affil[1]{\orgdiv{Laboratory of Engineering Thermodynamics (LTD)}, \orgname{RPTU Kaiserslautern}, \orgaddress{\city{Kaiserslautern}, \postcode{67663}, \country{Germany}}}

\abstract{
	Entropy scaling is a powerful technique that has been used for predicting transport properties of pure components over a wide range of states. 
	However, modeling mixture diffusion coefficients by entropy scaling is an unresolved task. 
	We tackle this issue and present an entropy scaling framework for predicting mixture self-diffusion coefficients as well as mutual diffusion coefficients in a thermodynamically consistent way. 
	The predictions of the mixture diffusion coefficients are made based on information on the self-diffusion coefficients of the pure components and the infinite-dilution diffusion  coefficients.
	This is accomplished using information on the entropy of the mixture, which is taken here from molecular-based equations of state.
	Examples for the application of the entropy scaling framework for the prediction of diffusion coefficients in mixtures illustrate its performance. 
	It enables predictions over a wide range of temperatures and pressures including gaseous, liquid, supercritical, and metastable states -- also for strongly non-ideal mixtures.
}


\maketitle

\section*{Introduction}

Diffusion in mixtures is important in many natural and technical processes. 
Numerical values for diffusion coefficients are needed in many applications, e.g., for the design of fluid separation processes, reactors, and combustion processes. 
However, experimental data on diffusion coefficients are notoriously scarce, so that reliable models for their prediction are needed. 
We introduce a model type that is based on entropy scaling and enables predictions that were previously infeasible.

\begin{figure}[h]
	\centering
	\includegraphics[width=8cm]{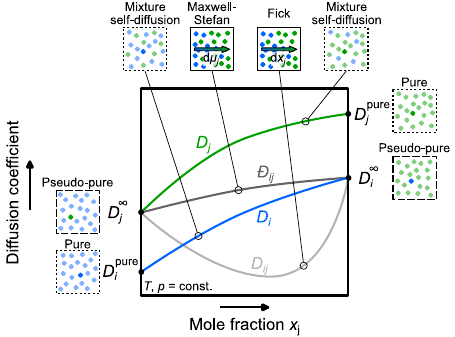}
	\caption{\textbf{Schematic representation of the different diffusion coefficients in a binary mixture as a function of the mole fraction.} Lines indicate the self-diffusion coefficient of component $i$ $D_i$ (blue) and component $j$ $D_j$ (green), the Maxwell-Stefan diffusion coefficient $\textit{\DH}_{ij}$ (dark grey), and the Fickian diffusion coefficient $D_{ij}$ (light grey). The limiting cases $D_i^{\rm pure}$, $D_j^{\rm pure}$ (both pure cases) and $D_i^{\infty}$, $D_j^{\infty}$ (both pseudo-pure cases) are also depicted.}
	\label{fig:scheme}
\end{figure}

In diffusion, two phenomena are distinguished: self-diffusion (a.k.a. tracer diffusion) and mutual (transport) diffusion. 
Self-diffusion and the corresponding self-diffusion coefficient $D_i$ of a component $i$ describes the Brownian movement of individual particles and is defined for pure components and mixtures. 
In contrast, mutual diffusion is only defined in mixtures and describes the motion of particle collectives of the different components resulting in macroscopic mass transfer. 
Obviously, self-diffusion and mutual diffusion are closely related, but there exist no generally applicable relation that connects self-diffusion coefficients and mutual diffusion coefficients.
The different diffusion coefficients are schematically depicted in Fig.~\ref{fig:scheme} for a binary mixture.

For describing mutual diffusion \cite{wesselingh_mass_2000}, there are two common frameworks: those of Fick and Maxwell-Stefan. 
The corresponding diffusion coefficients are the Fickian diffusion coefficients $D_{ij}$ (related to the concentration gradient as driving force) and the Maxwell-Stefan diffusion coefficients $\textit{\DH}_{ij}$ (related to the chemical potential gradient as driving force). 
The two frameworks are thermodynamically consistent representations and can be transformed into each other.
For a binary mixture, these diffusion coefficients are related by
\begin{equation}\label{eq:MS-Fick}
	D_{ij} = \textit{\DH}_{ij} \varGamma_{ij},
\end{equation}
where $\varGamma_{ij}$ is the thermodynamic factor, which is defined by 
\begin{equation}
	\varGamma_{ij} = \frac{x_i x_j}{{\rm R}T} \left( \frac{\partial^2 G}{\partial x_i^2} \right)_{T,p,n_{j \neq i}},
\end{equation}
where $x_i$ and $x_j$ are the mole fractions of components $i$ and $j$, respectively, R is the universal gas constant, and $G$ is the Gibbs energy of the mixture.

The two coefficients $D_{ij}$ and $\textit{\DH}_{ij}$ become equal if the thermodynamic factor is unity, which is the case in the infinite dilution limit (and for ideal mixtures). 
Furthermore, the self-diffusion coefficient and the mutual diffusion coefficients are related in the infinite dilution limit (cf. Fig. \ref{fig:scheme}). 
Thus, the following relations apply for the infinite dilution limit:
\begin{align}
	x_i \rightarrow 0:\quad & \textit{\DH}_{ij} = D_{ij} = D_i = D_i^{\infty} \label{eq:D_lim_1} \\
	x_j \rightarrow 0:\quad & \textit{\DH}_{ij} = D_{ij} = D_j = D_j^{\infty} \label{eq:D_lim_2} 
\end{align}
where $D_i^{\infty}$ is the diffusion coefficient of component $i$ infinitely diluted in component $j$, $D_j^{\infty}$ the diffusion coefficient of component $j$ infinitely diluted in component $i$.
Modeling the different diffusion coefficients in a mixture in a consistent way is a challenging task. 
In this work, we propose a methodology that provides such a framework.

Physical models for predicting mixture diffusion coefficients at gaseous states are known and established for a long time within kinetic gas theory \cite{elliott_properties_2023}.
The prediction of mixture diffusion coefficients at states where significant intermolecular interactions are present, on the other hand, is still an unresolved problem.
Einstein has proposed a method for estimating infinite-dilution diffusion  coefficients in liquids \cite{einstein_uber_1905}, for which several modifications exist today \cite{elliott_properties_2023,evans_improving_2018,zmpitas_modified_2021}.
For estimating the concentration dependence of mutual diffusion coefficients in mixtures, several empirical models have been proposed, e.g., the Vignes model and the generalized Darken model \cite{vignes_diffusion_1966,darken_diffusion_1948,liu_predictive_2011}. 
However, these models often fail for strongly non-ideal mixtures (see Suppl. Note 8).

In recent years, entropy scaling has received significant attention for modeling transport properties \cite{lotgering-lin_group_2015,bell_excess-entropy_2020,dyre_perspective_2018}.
It is based on the discovery of Rosenfeld \cite{rosenfeld_relation_1977,rosenfeld_quasi-universal_1999}, that dynamic properties (i.e. viscosity, thermal conductivity, and self-diffusion coefficient) of pure components, when properly scaled by the density and temperature, are a monovariate function of the configurational entropy (sometimes also referred to as 'residual entropy' or 'excess entropy').
This scaling behavior is physically based and related to isomorph theory \cite{gnan_pressure-energy_2009,dyre_perspective_2018}.
After being re-discovered by Dyre in a seminal review \cite{dyre_perspective_2018}, entropy scaling has become a popular approach in the last ten years. 
For modeling transport properties, the entropy scaling principle has been cast into many executable models using several different modified scaling approaches \cite{lotgering-lin_group_2015,bell_probing_2019,bell_modified_2019,schmitt_entropy_2024,dehlouz_entropy_2022,hopp_self-diffusion_2018}.
For obtaining the entropy at a desired state point (e.g., given by $T, p$), usually an equation of state (EOS) is used.
Entropy scaling can be favorably combined with molecular-based EOS\cite{nezbeda_molecular-based_2020}, which enables predictions beyond the available data \cite{schmitt_entropy_2024,wingertszahn_measurement_2023}.
Entropy scaling is well established for predicting the viscosity and thermal conductivity of mixtures, as for example demonstrated by Gross and co-workers \cite{lotgering-lin_group_2015,hopp_thermal_2017} and Bell and co-workers \cite{yang_entropy_2021,yang_entropy_2021-1} in recent years.
The corresponding models are based on combination and mixing rules and often enable a reliable prediction of mixture viscosities and thermal conductivities \cite{lotgering-lin_pure_2018,yang_entropy_2021,schmitt_entropy_2024}.
However, entropy scaling for mixture diffusion coefficients -- as depicted in Fig. \ref{fig:scheme} -- has not yet been developed.
So far, only the pure component limiting cases for the self-diffusion coefficients $D_1^{\rm pure}$ and $D_2^{\rm pure}$ (cf. Fig. \ref{fig:scheme}) can be described by entropy scaling models from the literature\cite{vaz_universal_2012,dehlouz_entropy_2022,hopp_self-diffusion_2018,bell_modified_2019,schmitt_entropy_2024}.
There are few molecular simulation studies of self-diffusion coefficients in mixtures available in the literature \cite{dzugutov_universal_1996,samanta_universal_2001,mittal_confinement_2007,pond_composition_2009,krekelberg_generalized_2009,carmer_enhancing_2012,bell_excess-entropy_2020,fertig_transport_2022} in which also the entropy scaling behavior of the data was considered.
They evaluate either simple model systems or special cases like metallic fluids. 
They show that self-diffusion coefficients of such fluids can follow a universal monovariate behavior, but the authors do not provide models for describing the self-diffusion coefficients.
Specifically, a quasi-universal scaling law for mixture diffusion coefficients has been proposed by Bell and Dyre \cite{bell_excess-entropy_2020} -- however, being limited to self-diffusion.
Truskett and co-workers \cite{mittal_confinement_2007,pond_composition_2009,krekelberg_generalized_2009,carmer_enhancing_2012} have studied the monovariate scaling behavior based on computer experiment data. 
Similarly, Fertig et al.\cite{fertig_transport_2022} have demonstrated that elements of monovariate scaling are also present in model mixture diffusion coefficient data. 
However, in none of these studies a generally applicable modeling framework for consistently describing the different diffusion coefficients in mixtures has been developed, so that this is still an unresolved issue.

In this work, we propose an entropy scaling model for predicting mixture diffusion coefficients, namely the self-diffusion coefficients as well as the (Fickian and MS) mutual diffusion coefficients, without any adjustable mixture parameters.
The physical framework we have developed for predicting diffusion coefficients in mixtures can be characterized as follows: 
(1) It can be applied in the entire fluid region, i.e. it covers gases, liquids, and supercritical fluids, phase equilibria, and even metastable states. 
(2) It describes both self-diffusion and mutual diffusion in a consistent way. 
(3) It comprises the dependence of the diffusion coefficients on temperature, pressure, and composition in the entire fluid region. 
The applicability of the model is demonstrated for binary mixtures in this work. 

The approach developed in this work is based on three central ideas and concepts:
(1) infinite-dilution diffusion  coefficients are treated as pseudo-pure components that exhibit a monovariate scaling behavior, which can be treated by classical entropy scaling \cite{schmitt_entropy_2024}. 
(2) Therefore, the mixture diffusion coefficient limiting cases, i.e. the pure component and pseudo-pure component self-diffusion coefficients (cf. Fig.~\ref{fig:scheme}), are modeled as functions of the entropy.
This requires at least one reference data point for each limiting case. 
(3) Based on the information of the limiting cases, the concentration dependence of the diffusion coefficients $D_i$, $D_j$, $\textit{\DH}_{ij}$, and $D_{ij}$ are predicted using combination and mixing rules. 
(4) It correctly captures the limits of the different diffusion coefficients to the pure component and infinite dilution.
We demonstrate that these predictions can succeed without any adjustable mixture parameter. 
The performance of the approach is demonstrated using model fluids as well as real substance systems. 
In this work, we employ molecular-based EOS models. 
Yet, also other EOS types such as multiparameter \cite{span_multiparameter_2000} or cubic EOS \cite{kontogeorgis_equation_1996} could be used.

\section*{Results}

First, we demonstrate the applicability of a monovariate scaling to the infinite dilution self-diffusion coefficients by treating $D_i^\infty$ as a pseudo-pure component property. 
This enables the prediction of $D_i^\infty$ to practically all fluid states based on very little data. 
The monovariate scaling of the infinite-dilution diffusion  coefficients uncovered in this work provides the basis for the modeling of the different mixture diffusion coefficients shown in the second part of this section.
The applicability of the modeling approach is demonstrated in the second part of this section, where model predictions for the self-diffusion as well as mutual diffusion coefficients are compared to reference data for model and real substance systems.

\subsection*{Infinite-Dilution Diffusion Coefficients}

\begin{figure}[h!]
	\centering
	\includegraphics[width=6cm]{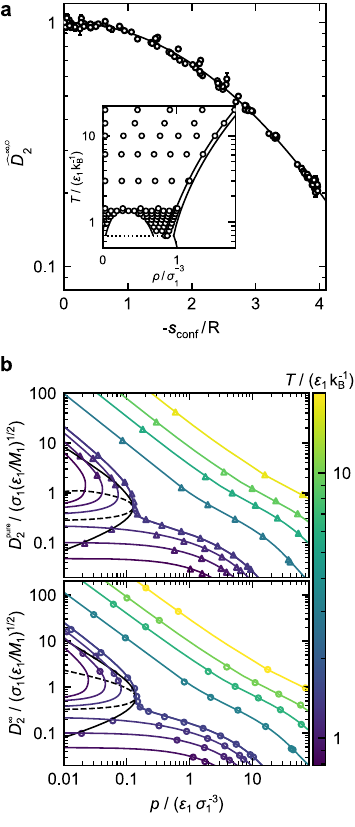}
	\caption{Entropy scaling of diffusion coefficients in a Lennard-Jones mixture with $\sigma_2 = \sigma_1$, $\varepsilon_2 = 0.9\,\varepsilon_1$, and $\varepsilon_{12} = 1.2\,\sqrt{\varepsilon_1\varepsilon_2}$. a) Scaled infinite-dilution diffusion  coefficient of component 2 $\widehat{D}^{\infty,\circ}_{2}$ as a function the reduced configurational entropy $s_{\rm conf}/{\rm R}$. The line indicates the entropy scaling model. Symbols are MD simulation data from this work. The inset shows the simulation state points in the temperature-density phase diagram of component 1. Therein, solid lines indicate the phase envelopes from Refs. \cite{stephan_thermophysical_2019,schultz_comprehensive_2018}. b) Self-diffusion coefficient of component 2 $D_2^{\rm pure}$ (top) and infinite-dilution diffusion  coefficient of component 2 $D^{\infty}_{2}$ (bottom) as function of the pressure. Lines are the entropy scaling model and symbols are simulation results from Ref. \cite{schmitt_entropy_2024} ($D_2^{\rm pure}$) and from this work ($D^{\infty}_{2}$). The black solid line indicates the vapor-liquid equilibrium and the black dashed line the corresponding spinodal. Colors indicate the temperature. Source data are provided as a Source Data file.}
	\label{fig:infdilution}
\end{figure}

Fig.~\ref{fig:infdilution}a demonstrates the monovariate scaling behavior for infinite-dilution diffusion  coefficients $\widehat{D}_{2}^{\infty,\circ}$ for a binary Lennard-Jones mixture.
Results are shown for the scaled self-diffusion coefficient of component 2 at infinite dilution $\widehat{D}_{2}^{\infty,\circ}$; in our notation, the hat $\widehat{\space}$ and $^\circ$ refer to specific parts of the scaling, see Methods.
The data collapse to a monovariate curve, which is impressive considering the fact that a large range of states was studied, cf. Fig. \ref{fig:infdilution}a-inset. 
The quality of the scaling of $\widehat{D}_{2}^{\infty,\circ}$ (i.e. how well the data can be described by a monovariate function) is essentially the same as that found for the pure component diffusion coefficient $D_2$ (see Suppl. Note 5 for quantification). 
This supports the picture introduced in this work that $D_{2}^{\infty}$ can be considered as a pseudo-pure component property.
This picture is physically meaningful considering the fact that in both cases, the mobility of a single particle in a homogeneous environment is described.
The Chapman-Enskog diffusion coefficient agrees well with the low-density simulation results -- as indicated by the convergence of $\widehat{D}^{\infty,\circ}_{2} \rightarrow 1$ for $s_{\rm conf} / {\rm R} \rightarrow 0$.

\begin{figure*}[h!]
	\centering
	\includegraphics[width=\textwidth]{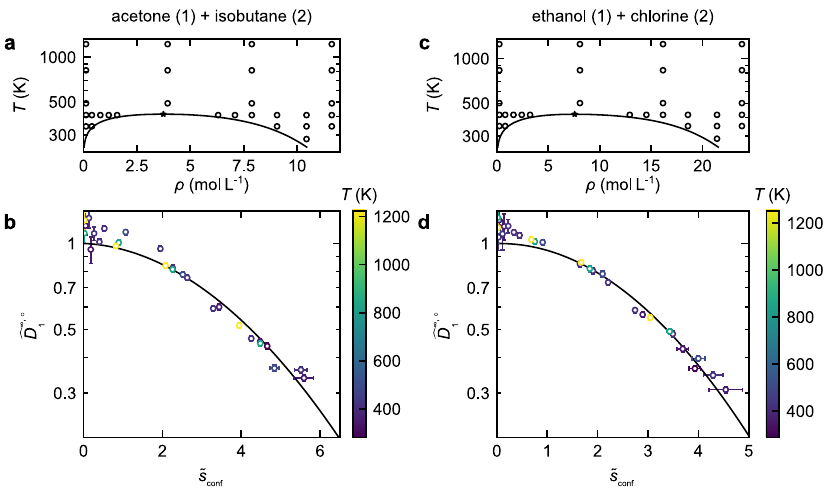}
	\caption{Scaling behavior of the infinite-dilution diffusion  coefficients in the system acetone (1) + isobutane (2) (left) and ethanol (1) + chlorine (2). a,c) Studied state points (symbols) in the temperature-density phase diagram of the solvents. The line indicates the vapor-liquid equilibrium and the star the critical point as calculated from the PC-SAFT EOS \cite{gross_perturbed-chain_2001}. b,d) Scaled infinite-dilution diffusion  coefficient of the solutes $\widehat{D}_{\rm 1}^{\infty,\circ}$ as a function of the reduced configurational entropy $\tilde{s}_{\rm conf}$. The symbols are the simulation results (color indicates the temperature) and the line the entropy scaling model (fitted to the simulation results). The error bars represent the uncertainties of the molecular simulation results. Source data are provided as a Source Data file.}
	\label{fig:real_scaling}
\end{figure*}

\reffig{fig:real_scaling} shows the simulation results for the infinite-dilution diffusion  coefficients in the systems acetone + isobutane and ethanol + chlorine.
The simulations comprise states in the gas, liquid, and supercritical regions.
Details on the simulations are provided in Suppl. Note 2.
For both systems, the infinite-dilution diffusion  coefficient shows a monovariate behavior with respect to the configurational entropy confirming the results for the Lennard-Jones model system.
Significant deviations from the monovariate behavior are observed for the real systems only for $\tilde{s}_{\rm conf}$, i.e. for low-density state points.
This might be due to larger uncertainties of the simulations in this region.

In Suppl. Note 6), the results for two more binary Lennard-Jones systems and a third real substance system are presented that support the findings discussed here.
For the Lennard-Jones model systems, the "global" parameters from Ref. \cite{schmitt_entropy_2024} were sufficient for describing $\widehat{D}_2^{\infty,\circ}$.
This finding was not expected and demonstrates the broad applicability of the universal parameters $g_1$ and $g_2$ in the correlation that were determined in Ref. \cite{schmitt_entropy_2024}.

From the models for the pure component self-diffusion coefficient $\widehat{D}_2^{\rm pure}(\tilde{s}_{\rm conf})$ \cite{schmitt_entropy_2024} and the infinite-dilution diffusion  coefficient $\widehat{D}_2^{\infty}(\tilde{s}_{\rm conf})$, both $D_2^{\rm pure}$ and $D_{2}^{\infty}$ can be predicted in a wide range of states. 
For the Lennard-Jones systems, the configurational entropy was taken from the Kolafa-Nezbeda EOS\cite{kolafa_lennard-jones_1994}, 
which is known to give an accurate the description of Lennard-Jones fluids \cite{antolovic_phase_2023,stephan_review_2020}. 
We used the mixture implementation for the Kolafa-Nezbeda EOS from Refs. \cite{antolovic_phase_2023,schmitt_entropy_2024}.
Fig. \ref{fig:infdilution}b shows the results of the entropy scaling models for $D_2$ and $D_{2}^{\infty}$ in comparison to the MD results. 
Both methods are in good agreement, which is a result of the good performance of the EOS as well as of the fact that both properties show a monovariate relation with respect to the configurational entropy. 

Hence, reliable analytical models for the limiting cases $D_1$, $D_2$, $D_1^{\infty}$, and $D_2^{\infty}$ (cf. Fig.~\ref{fig:scheme}) are now available that can be applied in all fluid regions. 
They are the basis for the next step, the prediction of $D_1$, $D_2$, $\textit{\DH}_{12}$, and $D_{12}$ at arbitrary compositions in the mixture.

\subsection*{Diffusion Coefficients in Mixtures}

\begin{figure*}[h!]
	\centering
	\includegraphics[width=\textwidth]{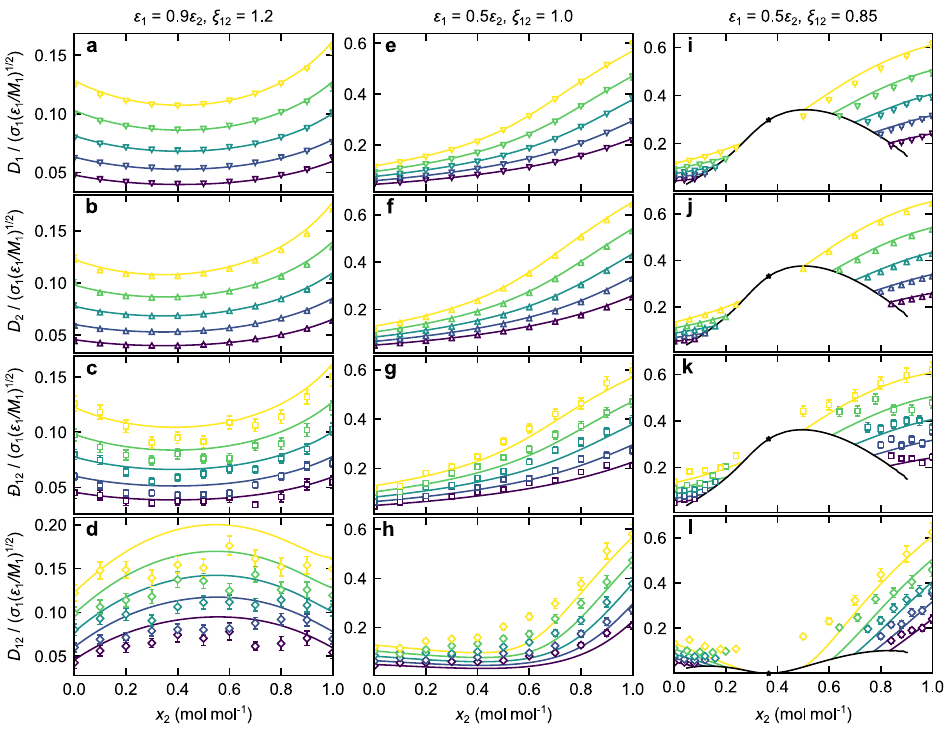}
	\caption{Diffusion coefficients in three binary Lennard-Jones mixture with $\sigma_2 = \sigma_1$ as a function of the mole fraction $x_2$ at $p = 0.13\,\sigma^3_1\varepsilon_1^{-1}$ (a-d) and $p = 0.26\,\sigma^3_1\varepsilon_1^{-1}$ (e-l). Energy parameters $\varepsilon_1, \varepsilon_2$ and mixing parameter $\xi_{12}$ are given for each column. a,e,i) Self-diffusion coefficient of component 1 $D_1$; b,f,j) Self-diffusion coefficient of component 2 $D_2$; c,g,k) Maxwell-Stefan diffusion coefficient $\textit{\DH}_{12}$; d,h,l) Fickian diffusion coefficient $D_{12}$. Lines are predictions from the entropy scaling framework. Symbols are simulation results from Ref. \cite{fertig_influence_2023}. The entropy scaling framework was used in combination with the Kolafa-Nezbeda EOS\cite{kolafa_lennard-jones_1994}. The colors indicate the temperature $T \in \{ 0.79,0.855,0.92,0.985,1.05 \}\,\KB{}\varepsilon_1^{-1}$ (from blue to yellow). The black lines indicate the liquid-liquid equilibrium and the star the critical point (only i-j). The error bars represent the simulation uncertainty given in Ref. \cite{fertig_influence_2023}. Source data are provided as a Source Data file.}
	\label{fig:LJ_mix}
\end{figure*}

The model proposed in this work (see Methods) is able to predict all diffusion coefficients in binary mixtures.
Fig. \ref{fig:LJ_mix} shows the model predictions for all four diffusion coefficients ($D_1$, $D_2$, $\textit{\DH}_{12}$, $D_{12}$) of three Lennard-Jones systems for the entire composition range and different temperatures.
For comparison, simulation data from Ref. \cite{fertig_influence_2023} are used.
The predictions from the entropy scaling framework and the computer experiment results agree well for all investigated systems and all four diffusion coefficients (see Fig.~\ref{fig:LJ_mix}). 

For both self-diffusion coefficients, for which the simulation results have a smaller statistical uncertainty than for the mutual diffusion coefficients, the entropy scaling model describes most simulation data within their uncertainty, despite the strong non-ideality that leads to extrema at about $x_2 \approx 0.4\,{\rm mol\,mol^{-1}}$  (a-d in Fig. \ref{fig:LJ_mix}) or even to the occurrence of a liquid-liquid equilibrium gap (i-l in Fig. \ref{fig:LJ_mix}) -- the phase behavior of these model systems was comprehensively studied in Ref. \cite{stephan_influence_2020}.
The simulation data for $\textit{\DH}_{12}$ scatter more and have larger error bars than those for $D_1$ and $D_2$. 
The entropy scaling framework predicts the $\textit{\DH}_{12}$ simulation data very well in the entire composition range.
In particular, the strongly non-ideal behavior is captured accurately by the entropy scaling framework.
We have compared the entropy scaling framework presented in this work to results from the established Vignes and Darken model (see Suppl. Note 8).
The entropy scaling model outperforms the empirical models. 
Also the temperature and pressure dependency of the diffusion coefficients are very well described by the model (cf. Fig. \ref{fig:LJ_mix} and Suppl. Note 7, respectively).
For the predictions of the Fickian diffusion coefficient, the trends are correctly predicted by the entropy scaling predictions, but deviations are observed, cf. Fig. \ref{fig:LJ_mix}d,h,l. 
The deviations between simulation results and entropy scaling are larger for the Fickian diffusion coefficient, which can be attributed to the inclusion of the thermodynamic factor adding additional complexity to the model predictions (and the MD sampling \cite{fertig_transport_2022}).
In contrast to the other three diffusion coefficients ($D_1$, $D_2$, $\textit{\DH}_{12}$), the Fickian diffusion coefficient $D_{12}$ exhibits the opposite extremum behavior, which is a result of the underlying thermodynamic factor behavior.
Due to the coupling of entropy scaling and EOS, the model is inherently consistent and can be applied over a wide range of states. 
As an example, Fig.~\ref{fig:LJ_mix} (i-j) includes a system with weak cross-interactions resulting in a liquid-liquid equilibrium at the considered conditions. 
The entropy scaling model proposed in this work is able to predict not only the different diffusion coefficients in the bulk phases, but also describes the diffusion coefficients of the coexisting phases of the liquid-liquid equilibrium.
Also, at the upper critical solution point, the thermodynamic factor becomes zero and so does the Fickian diffusion coefficient ($x_2 \approx 0.38 \, {\rm mol\,mol^{-1}}$).
This is as expected for the Fickian diffusion coefficient at a critical point \cite{antolovic_phase_2023} and correctly captured by the model.

\begin{figure}[h!]
	\centering
	\includegraphics[width=6cm]{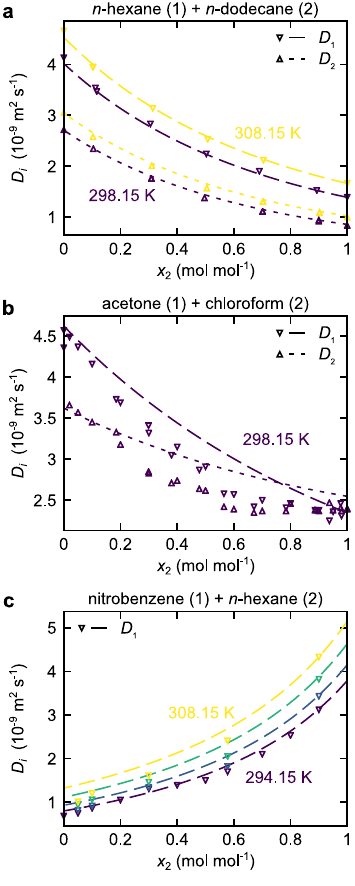}
	\caption{
		Self-diffusion coefficients $D_i$ of real substance mixtures predicted by the entropy scaling model as a function of the mole fraction $x_2$ at $p = 0.1\,{\rm MPa}$. 
		Symbols are experimental data from the literature and lines are model predictions.
		a) Mixture $n$-hexane (1) + $n$-dodecane (2); experimental data from Ref. \cite{shieh_transport_1969}.
		b) Mixture acetone (1) + chloroform (2); experimental data from Ref. \cite{dagostino_prediction_2013}.
		c) Mixture nitrobenzene (1) + $n$-hexane (2); experimental data from Ref. \cite{dagostino_prediction_2011}.
		Source data are provided as a Source Data file.
	}
	\label{fig:real_self}
\end{figure}

\begin{figure}[h!]
	\centering
	\includegraphics[width=6cm]{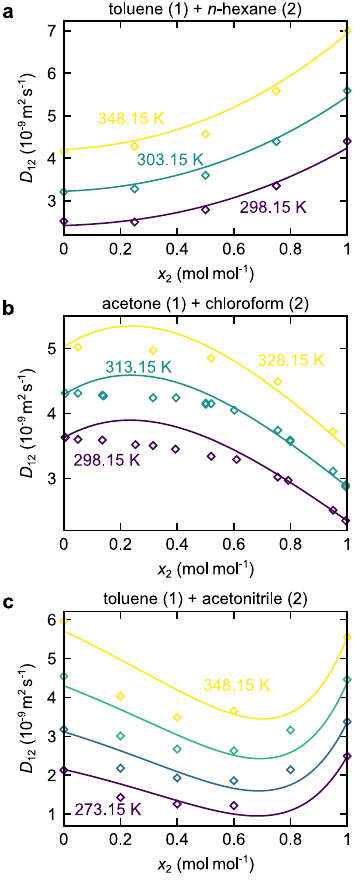}
	\caption{
		Fickian diffusion coefficients of real substance mixtures predicted by the entropy scaling model as a function of the mole fraction $x_2$ at $p = 0.1\,{\rm MPa}$. 
		Symbols are experimental data and lines are model predictions.
		a) Mixture toluene (1) + $n$-hexane (2); experimental data from Ref. \cite{chen_determination_1992}.
		b) Mixture acetone (1) + chloroform (2); experimental data from Refs. \cite{anderson_mutual_1958,tyn_temperature_1975}.
		c) Mixture toluene (1) + acetonitrile (2); experimental data from Ref. \cite{awan_transport_2001}.
		Source data are provided as a Source Data file.
	}
	\label{fig:real_mutual}
\end{figure}

Fig. \ref{fig:real_self} shows the results for the self-diffusion coefficients in three real substance systems $n$-hexane + $n$-dodecane, acetone + chloroform, and nitrobenzene + $n$-hexane.
Fig. \ref{fig:real_mutual} shows entropy scaling predictions for the Fickian diffusion coefficient in four real substance systems: toluene + $n$-hexane, acetone + chloroform, and toluene + acetonitrile.
For all systems, only very limited experimental data are available (this is common for practically all real mixtures), which is used for the validation of the predictions. 
The thermodynamic properties of the systems were modeled by the PC-SAFT EOS \cite{gross_perturbed-chain_2001} and the component-specific models for the pure substances were taken from the literature (see Suppl. Note 3 for details).
The Berthelot combination rule parameter $\xi_{12}$ for the systems acetone + chloroform and nitrobenzene + $n$-hexane were adjusted to match the respective vapor-liquid equilibria qualitatively.

The entropy scaling predictions of both self-diffusion coefficients and the experimental data are in good agreement. 
Only for the system acetone + chloroform, some deviations are observed. 
For the systems $n$-hexane + $n$-dodecane and nitrobenzene + $n$-hexane, the model predicts the experimental data very well; the mean relative deviations are $\overline{\updelta D_1} = 1.24\,\%$ and $\overline{\updelta D_2} = 2.91\,\%$ ($n$-hexane + $n$-dodecane) and $\overline{\updelta D_1} = 4.96\,\%$ (nitrobenzene + n-hexane), respectively.
For the system $n$-hexane + $n$-dodecane, the two components strongly differ in their molar mass, which yields a non-linear behavior of $D_1(x_2)$ and $D_2(x_2)$.
This behavior is predicted well by the framework.
Also for the system nitrobenzene + hexane, the non-ideality of the diffusion coefficients is well captured by the model. 
For the system acetone + chloroform, the mean relative deviations are $\overline{\updelta D_1} = 8.58\,\%$ and $\overline{\updelta D_2} = 9.69\,\%$.

Fig. \ref{fig:real_mutual} shows the results for the Fickian diffusion coefficient $D_{12}$ in three systems, namely toluene + $n$-hexane, acetone + chloroform, and toluene + acetonitrile.
All results are at ambient pressure, but at different temperatures and shown as a function of the mole fraction. 
The Fickian diffusion coefficient increases with increasing temperature.
The entropy scaling predictions are in good agreement with the experimental data ($\overline{\updelta D_{12}} = 3.02 \, \%$ for toluene + $n$-hexane, $\overline{\updelta D_{12}} = 3.67 \, \%$ for acetone + chloroform, and $\overline{\updelta D_{12}} = 10.27 \, \%$ for toluene + acetonitrile).
For the system toluene + $n$-hexane, the entropy scaling model yields good agreement. 
For the systems toluene + acetonitrile and acetone + chloroform, some deviations are observed - especially in the vicinity of the extremum values. 

\begin{figure}[h]
	\centering
	\includegraphics[width=6cm]{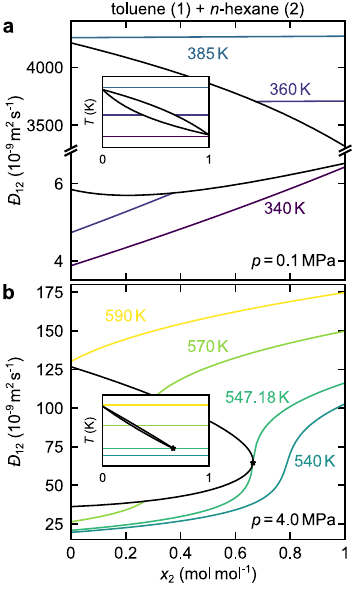}
	\caption{Maxwell-Stefan diffusion coefficient $\textit{\DH}_{12}$ in the mixture toluene (1) + $n$-hexane (2) as a function of the mole fraction of $n$-hexane $x_{2}$ at $p = 0.1\,{\rm MPa}$ (a) and $p = 4\,{\rm MPa}$ (b). Black lines: Vapor-liquid equilibrium; colored lines: Isotherms. Insets show the $T-x$ vapor liquid-equilibrium phase diagram at the respective pressure. Source data are provided as a Source Data file.}
	\label{fig:real_extrapolation}
\end{figure}

Fig. \ref{fig:real_extrapolation} shows the predictions for the Maxwell-Stefan diffusion coefficient phase diagrams of the system toluene + $n$-hexane at two pressures: $p=0.1\,{\rm MPa}$ (a) and $p = 4\,{\rm MPa}$ (b).
The diffusion coefficients of the coexisting phases can be obtained in a straightforward manner by the framework. 
Additional results (including supercritical, metastable, and unstable states) are presented in Suppl. Note 10. 
At $p=0.1\,{\rm MPa}$, both components are subcritical (see inset in Fig. \ref{fig:real_extrapolation} a).
The Maxwell-Stefan diffusion coefficient is shown for three isotherms (340, 360, 385 K).
As shown in the inset (see Fig. \ref{fig:real_extrapolation} a), one isotherm is entirely in the liquid phase (340 K), one isotherm passes through the vapor-liquid equilibrium (360 K), and one isotherm is entirely in the gas phase (385 K).
The Maxwell-Stefan diffusion coefficient in the gas phase is significantly larger than that in the liquid phase.
In the liquid phase, the diffusion coefficient strongly depends on the composition.
In the gas phase, the Maxwell-Stefan diffusion coefficient exhibits only a minor dependency on the composition.
This is in line with the Chapman-Enskog theory\cite{elliott_properties_2023} -- which is inherently incorporated in our framework. 
At $p = 4\,{\rm MPa}$ (cf. Fig. \ref{fig:real_extrapolation} b), $n$-hexane is supercritical and a critical point exists at about $T = 547.18\,{\rm K}$ (see inset).
Four isotherms are plotted: One subcritical isotherm ($T = 540\,{\rm K}$), the critical isotherm ($T= 547.18\,{\rm K}$), one isotherm passing through the vapor-liquid equilibrium ($T = 570\,{\rm K}$), and one isotherm entirely in the gas/supercritical phase ($T = 590\,{\rm K}$).
The isotherm at $T = 540\,{\rm K}$ exhibits a transition between the two pseudo-pure component limiting cases -- from a liquid state at $x_2 = 0$ to a supercritical state at $x_2 = 1\;{\rm mol\,mol^{-1}}$. 
At the critical point, the diffusion coefficient exhibits a large gradient with respect to the composition.
Due to the thermodynamic consistency established by the EOS, the framework correctly predicts the Fickian diffusion coefficient to be zero at critical points (see \reffig{fig:LJ_mix}).
In Suppl. Note 10, it is shown that also diffusion coefficients of metastable and unstable states as well as at spinodal states can be described by the framework (which is for example relevant for nucleation). 
This significantly exceeds the capabilities of all presently available diffusion coefficient models. 

\section*{Discussion} 

In this work, a methodology for predicting diffusion coefficients in mixtures in a consistent manner for all fluid states is proposed.
This includes self-diffusion and mutual diffusion coefficients.
The approach combines several physical concepts, including the modified Rosenfeld scaling, the Chapman-Enskog theory, and a molecular-based EOS. 
The framework enables predictions for the mixture diffusion coefficients without any adjustable parameters based on the limiting cases of the pure components and pseudo-pure components (i.e. infinite dilution). 
Due to the coupling of entropy scaling and an EOS, the framework can consistently describe diffusion coefficients in mixtures in different phases (gas, liquid, supercritical, metastable) -- including coexisting phases such as vapor-liquid and liquid-liquid equilibria. 
Even predictions to regions in which no data are available are possible. 
The strong predictive capabilities are a result of the physical backbone of the framework. 
The fact that infinite-dilution diffusion  coefficients exhibit a monovariate scaling behavior was uncovered here for the first time.
It enables an efficient entropy scaling modeling of the limiting mixture diffusion coefficient cases based on very few data. 
Since the limiting case diffusion coefficients are monovariate functions with respect to the entropy, extrapolations beyond the temperature and pressure range where data are available are feasible. 
In the mixture, no general monovariate behavior retains. 
Nevertheless, the mixture entropy in combination with mixing and combination rules enables predictions of mixture diffusion coefficients, which was demonstrated here for model mixtures as well as real substance systems. 

The entropy scaling framework requires an accurate and robust EOS model that describes mixture thermodynamic properties reliably, i.e. the phase equilibria, the entropy, the thermodynamic factor, and the second virial coefficient. 
If such a model is available, the entropy scaling framework proposed in this work is a powerful tool. 
For future work, the extension to multi-component systems, e.g. Refs. \cite{safi_measurement_2007,safi_diffusion_2008,su_fick_2025}, would be interesting.

\section*{Methods} \label{sec:Methods}

The framework proposed in this work consists of two elements: 
(1) Treating the infinite dilution self-diffusion coefficient as a pseudo-pure component to obtain a monovariate scaling; 
(2) The application of mixing and combination rules for predicting the different diffusion coefficients in a mixture. 
Throughout this work, the configurational entropy $s_{\rm conf}(T,\rho,\underline{x})$ is calculated from molecular-based EOS as the derivative of the configurational Helmholtz energy $a_{\rm conf}$ with respect to temperature $T$ at constant volume $v$ and composition $\underline{x}$, i.e.
\begin{equation}
	s_{\rm conf} = - \left( \frac{\partial a_{\rm conf}}{\partial T} \right)_{v,\underline{x}}.
\end{equation}
Therein, the configurational Helmholtz energy $a_{\rm conf}$ is defined as $a_{\rm conf} = a - a_{\rm ideal}$ with $a_{\rm ideal}$ being the Helmholtz energy of the ideal gas.
We use a reduced configurational entropy defined as
\begin{equation}\label{eq:s_reduced}
	\tilde{s}_{\rm conf} = -s_{\rm conf}/({\rm R} m),
\end{equation}
where $m$ is the segment parameter of molecular-based EOS.

\subsection*{Entropy scaling of infinite-dilution diffusion coefficients}

The infinite-dilution diffusion coefficient $D_{i}^{\infty}(T,p)$ is treated as a pseudo-pure component property such that the corresponding scaled property $\widehat{D}_{i}^{\infty,\circ}(\tilde{s}_{\rm conf})$ exhibits a monovariate function with respect to the reduced configurational entropy $\tilde{s}_{\rm conf}$ (cf. \refequ{eq:s_reduced}).
The modeling methodology is described in the following.
There are many scaling approaches available in the literature that aim at establishing a monovariate behavior of transport coefficients. 
Here, we use the scaling described in Ref. \cite{schmitt_entropy_2024}. 
The pure component scaling described in Ref. \cite{schmitt_entropy_2024} (which follows Refs. \cite{rosenfeld_quasi-universal_1999,bell_modified_2019}) is adapted here to model the pseudo-pure component diffusion coefficients at infinite dilution. 
Thus, we propose that the infinite-dilution diffusion  coefficient is transformed using
\begin{equation}\label{eq:Mod_Rosenfeld_scaling}
	D_{i}^{\infty,\circ} = D_{i}^{\infty} \frac{\rho_{\rm N}^{1/3}}{\sqrt{{\rm R} T / M_{\rm CE}}}\left( \frac{-s_{\rm conf}}{{\rm R}} \right)^{2/3},
\end{equation}
where $\rho_{\rm N}$ is the number density of the solvent at given $T$ and $p$. 
The reference mass $M_{\rm CE}$ is adopted from the Chapman-Enskog (CE) theory as
\begin{equation}
	M_{\rm CE} = \frac{2}{1/M_i + 1/M_j},
\end{equation}
where $M_i$ and $M_j$ are the molar masses of the pure components.
The modified Rosenfeld-scaled diffusion coefficient $D_{i}^{\infty,\circ}$ exhibits some scattering in the zero-density limit for $s_{\rm conf} \rightarrow 0$.
Therefore, the scaling is further modified using the Chapman-Enskog diffusion coefficient $D_{{\rm CE},i}^{\infty,\circ}$, which is computed using a Lennard-Jones kernel (see Suppl. Note 1 for details).
Finally, the CE-scaled infinite-dilution diffusion  coefficient $\widehat{D}^{\infty,\circ}_{i}$ is described by a split between the low-density (LD) and the high-density (HD) region as
\begin{equation}\label{eq:CEscaled_D}
	\widehat{D}^{\infty,\circ}_{i} = \underbrace{ \frac{D^{\infty,\circ}_{i}}{D^{\infty,\circ}_{{\rm CE},i}} }_{\rm LD} W(\tilde{s}_{\rm conf}) + \underbrace{ \frac{D^{\infty,\circ}_{i}}{\min \left( D^{\infty,\circ}_{{\rm CE},i} \right) }}_{\rm HD} (1-W(\tilde{s}_{\rm conf})),
\end{equation}
where $W$ is a smoothed transition function given by $W(\tilde{s}_{\rm conf}) = 1 / (1 + \exp(20(\tilde{s}_{\rm conf} - \tilde{s}_{\rm conf}^{\times})))$ with $\tilde{s}_{\rm conf}^{\times} = 0.5$.
The function $W$ establishes a smooth transition from the low-density region to the high-density region.
The choice of $\tilde{s}_{\rm conf}^{\times}$ is heuristic.
The resulting CE-scaled infinite-dilution diffusion  coefficient can be described by a simple continuous, monovariate function $\widehat{D}^{\infty,\circ}_{i} = \widehat{D}^{\infty,\circ}_{i}(\tilde{s}_{\rm conf})$.
Here, a function with two adjustable, system-dependent parameters $\alpha_{2}^{(D^\infty_i)}$ and $\alpha_{3}^{(D^\infty_i)}$ was used given as
\begin{equation}\label{eq:correlation}
	\ln \left( \widehat{D}^{\infty,\circ}_{i} \right) = \frac{\alpha_{2}^{(D^{\infty}_i)} (\tilde{s}_{\rm conf})^2  + \alpha_{3}^{(D^{\infty}_i)}(\tilde{s}_{\rm conf})^3}{1 + g_{1}^{(D)} \ln(\tilde{s}_{\rm conf}+ 1) + g_{2}^{(D)} \tilde{s}_{\rm conf}},
\end{equation}
where $g_1^{(D)} = 0.6632$ and $g_2^{(D)} = 9.4714$ are global parameters fitted to the self-diffusion coefficient of the Lennard-Jones fluid \cite{schmitt_entropy_2024}.
The system-dependent parameters $\alpha_{2}^{(D^{\infty}_i)}$ and $\alpha_{3}^{(D^{\infty}_i)}$ were fitted to reference data of the pseudo-pure component, i.e. the diffusion coefficients at infinite dilution.
This scaling approach was tested using model systems and real substance systems (see Results).
Therefore, MD simulations were carried out using the simulation engine ms2 \cite{fingerhut_ms2_2021}.

\subsection*{Predicting Diffusion Coefficients in Mixtures}

For predicting the diffusion coefficients $D_i(x_j)$, $D_j(x_j)$, $\textit{\DH}_{ij}(x_j)$, and $D_{ij}(x_j)$ in a mixture, the entropy scaling framework is extended using combining and mixing rules.
Diffusion coefficients in mixtures are predicted based on the limiting case models, i.e. the models for the self-diffusion coefficient of the pure components and the (infinite dilution) pseudo-pure component.
The self-diffusion coefficient $D_i$ in a mixture is calculated using the limiting case entropy scaling model of the pure component self-diffusion coefficient $D_i^{\rm pure}$ and the limiting case entropy scaling model of the infinite-dilution diffusion  coefficient $D_i^{\infty}$ in the solvent $j$.
The Maxwell-Stefan diffusion coefficient is calculated using both infinite-dilution diffusion  coefficient limiting case entropy scaling models $D_i^{\infty}$ and $D_j^{\infty}$.
Therefore, the scaled mixture diffusion coefficient $\widehat{\Lambda}^{\circ} \in \{ \widehat{D}_{i}^{\circ}, \widehat{D}_{j}^{\circ}, \widehat{\textit{\DH}}_{ij}^{\circ} \}$ is computed as
\begin{equation}\label{eq:correlation_mix}
	\ln \left( \widehat{\Lambda}^{\circ} \right) = \frac{\alpha_{2}^{(\Lambda)} (\tilde{s}_{\rm conf})^2  + \alpha_{3}^{(\Lambda)}(\tilde{s}_{\rm conf})^3}{1 + g_{1}^{(D)} \ln(\tilde{s}_{\rm conf}+ 1) + g_{2}^{(D)} \tilde{s}_{\rm conf}}, 
\end{equation}
where $\tilde{s}_{\rm conf}$ is the scaled configurational entropy of the mixture, i.e. $\tilde{s}_{\rm conf} = s_{\rm conf} / ({\rm R} m_{\rm mix})$ with $m_{\rm mix} = x_i m_i + x_j m_j$.
The parameters $\alpha_{k}^{(\Lambda)}$ (with $k \in \{ 2,3 \}$) are calculated using linear mixing rules
\begin{align}
	\alpha_{k}^{(D_i)}       & = x_i \alpha_k^{(D_i^{\rm pure})} + x_j \alpha_k^{(D_{i}^{\infty})}, \\
	\alpha_{k}^{(D_j)}       & = x_i \alpha_k^{(D_{j}^{\infty})} + x_j \alpha_k^{(D_j^{\rm pure})}, \\
	\alpha_{k}^{(\textit{\DH}_{ij})} & = x_i \alpha_k^{(D_{j}^{\infty})} + x_j \alpha_k^{(D_{i}^{\infty})},
\end{align}
for the self-diffusion coefficients of components $i$, the self-diffusion coefficient of component $j$, and the Maxwell-Stefan diffusion coefficient, respectively.

The final (unscaled) diffusion coefficient $\Lambda \in \{ D_i, D_j, \textit{\DH}_{ij} \}$ is calculated as
\begin{equation}\label{eq:Rosenfeld_mix}
	\Lambda = \frac{\widehat{\Lambda}^{\circ}}{ \frac{W(\tilde{s}_{\rm conf}) }{ \Lambda^{\circ}_{\rm CE} } + \frac{ 1-W(\tilde{s}_{\rm conf}) }{ \min \left( \Lambda^{\circ}_{\rm CE} \right) } } \cdot  \sqrt{\frac{{\rm R}T}{M_{\rm ref}^{\Lambda}}} \frac{1}{\rho_{\rm N}^{1/3}} \left( \frac{-s_{\rm conf}}{{\rm R}} \right)^{-2/3},
\end{equation}
where $\Lambda^{\circ}_{\rm CE}$ is the Chapman-Enskog diffusion coefficient of the mixture, $\rho_{\rm N}$ is the number density of the mixture, and $M_{\rm ref}^\Lambda$ is the reference mass of the mixture, which is calculated as
\begin{align}
	M_{\rm ref}^{D_i} &= x_i M_i + x_j M_{\rm CE}, 					\\
	M_{\rm ref}^{D_j} &= x_i M_{\rm CE} + x_j M_j, \quad {\rm and} \\
	M_{\rm ref}^{\textit{\DH}_{ij}} &= M_{\rm CE}
\end{align}
for the self-diffusion coefficients $D_i$ and $D_j$ and the Maxwell-Stefan diffusion coefficient $\textit{\DH}_{ij}$, respectively.
Details on the calculation of the Chapman-Enskog property of the mixture $\Lambda^{\circ}_{\rm CE}$, the reference mass $M_{\rm ref}$, and $\widehat{\Lambda}^\circ = \widehat{\Lambda}^\circ(s_{\rm conf}^{\rm mix},\alpha_{k,{\rm mix}})$ are given in Suppl. Note 1.

The Fickian diffusion coefficient is computed from the Maxwell-Stefan diffusion coefficient by the thermodynamic factor as predicted by the EOS, see \refequ{eq:MS-Fick}.
Thus, all required quantities are obtained using predictive combination rules and mixing rules and the EOS mixture model.
No adjustable mixture parameters are introduced.
However, for modeling the diffusion coefficients of a given binary system based on a given EOS model, adjustable parameters have to be determined for the four limiting cases, i.e. the two pure components and the two infinite dilution pseudo-pure components, cf. $\alpha$ parameters in \refequ{eq:correlation}. 
For predicting the diffusion coefficients in the mixture, no adjustable parameters are required. 
The non-ideality of the mixture is primarily taken into account by the underlying EOS via the predicted configurational mixture entropy.
The combination of the EOS and entropy scaling enables the consistent calculation of the diffusion coefficients ($D_i$, $D_j$, $\textit{\DH}_{ij}$, $D_{ij}$), homogeneous bulk properties ($pvT$, $s_{\rm conf}$, $\varGamma_{ij}$, etc.), and phase equilibria (e.g. vapor-liquid and liquid-liquid equilibria).

%
%
%

\section*{Acknowledgements}

We gratefully acknowledge funding from the European Union’s Horizon Europe research and innovation program under grant agreement no. 101137725 (BatCAT) and from the German Research Foundation (DFG) under grant 548115878.
The simulations were carried out on the HPC machine ELWE at the RHRZ under the grant RPTU-MTD and on the HPC machine MOGON at the NHR SW under the grant TU-MSG (supported by the Federal Ministry of Education and Research and the state governments).

\section*{Author Contribution Statement}

S. Schmitt: Data curation, Formal analysis, Methodology, Software (lead), Visualization, Writing – original draft. H. H.: Funding acquisition, Writing – review \& editing (support). S. Stephan: Conceptualization, Methodology, Supervision, Funding acquisition, Software (support), Writing – review \& editing (lead).

\section*{Competing Interests}

The authors declare no competing interests.

\end{document}


	
	\title[Entropy scaling for diffusion coefficients in fluid mixtures]{Entropy scaling for diffusion coefficients in fluid mixtures}
	\makeatletter
	\let\printtitle\@title
	\makeatother
	
	\renewcommand{\thesection}{Suppl. Note \arabic{section}}
	\setcounter{section}{0}
	
	\renewcommand{\thesubsection}{Suppl. Note \arabic{section}.\arabic{subsection}}
	\setcounter{section}{0}
	
	\renewcommand{\figurename}{\textbf{Suppl. Fig.}}
	\renewcommand{\thefigure}{\textbf{\arabic{figure}}}
	\setcounter{figure}{0}
	
	\renewcommand{\tablename}{Suppl. Table}
	\renewcommand{\thetable}{\arabic{table}}
	\setcounter{table}{0}
	
	\renewcommand{\theequation}{\arabic{equation}}
	\setcounter{equation}{0}
	
	\clearpage
	\setcounter{page}{1}
	
	\begin{titlepage}
		\centering
		\vspace{8cm}
		{\scshape\huge Supplementary Information\par}
		\vspace{4cm}
		{\Huge\bfseries \printtitle \par}
		\vspace{4cm}
		{\Large {Sebastian~Schmitt, Hans~Hasse, and Simon~Stephan} \par}
		\vspace{2cm}
		
		\large{Laboratory~of~Engineering~Thermodynamics~(LTD),\\
			RPTU~Kaiserslautern,~Kaiserslautern,~Germany }\\[2cm] 
		\large {simon.stephan@rptu.de}\\[1cm] 

		\vfill
		
		
	\end{titlepage}
	
	
	
	\section{Scaled Chapman-Enskog diffusion coefficients}
	
	\subsection{Infinite-dilution diffusion coefficient}
	
	The scaled Chapman-Enskog infinite-dilution diffusion coefficient $D_{{\rm CE},i}^{\infty,\circ}$ is only a function of the temperature and given as
	\begin{equation}\label{eq:D_CE_plus}
		D_{{\rm CE},i}^{\infty,\circ} = \frac{3}{8 \sqrt{\uppi}} \frac{1}{\sigma_{ij}^2 \Omega^{(1,1)}}	\left(T \left(\frac{{\rm d}B}{{\rm d}T}\right) + B \right)^{2/3},
	\end{equation}
	where $B$ is the second virial coefficient of the solvent at a given temperature (which is computed from the EOS model), $\sigma_{ij}$ is the cross-interaction Lennard-Jones size parameter, and $\Omega^{(1,1)}_{ij} = \Omega^{(1,1)}(T \KB \varepsilon_{ij}^{-1})$ is the collision integral for diffusion \cite{elliott_properties_2023,kim_high-accuracy_2014}.
	The Lennard-Jones parameters $\sigma_{ij}$ and $\varepsilon_{ij}$ are calculated according to the Lorentz-Berthelot combining rules \cite{lorentz_ueber_1881,berthelot_sur_1898} from the pure component interaction parameters as
	\begin{align}
		\sigma_{ij} &= \frac{\sigma_i + \sigma_j}{2}\quad {\rm and} \label{eq:lorentz} \\
		\varepsilon_{ij} &= \xi_{ij} \sqrt{\varepsilon_j\varepsilon_i}, \label{eq:berthelot}
	\end{align}
	where $\xi_{ij}$ is a state-independent mixture parameter to establishing a non-ideality in the system
	The Lennard-Jones parameters $\sigma_{i}$, $\sigma_{j}$, $\varepsilon_{i}$, and $\varepsilon_{j}$ of a given pure (possibly real) component are calculated by the corresponding states principle with the Lennard-Jones fluid as reference from the critical temperature $T_{{\rm c,}j}$ and the critical pressure $p_{{\rm c,}j}$ of the solvent as $\varepsilon_{j} = T_{{\rm c},j} \KB / 1.321$ and $\sigma_{j} = (\varepsilon_{j} 0.129 / p_{{\rm c},j})^{1/3}$.

	\subsection{Diffusion coefficients in mixtures}
	
	The mutual diffusion coefficient in gases does not depend on the composition for mixtures and the Fickian and the Maxwell-Stefan diffusion coefficients are equal, i.e. $D_{ij} = \DMS_{ij}$.
	The Chapman-Enskog self-diffusion coefficients in the mixture $D_{i,{\rm CE}}^{\circ}(\underline{x})$ and $D_{j,{\rm CE}}^{\circ}(\underline{x})$ are calculated according to Miller and Carman \cite{miller_self-diffusion_1961} as
	\begin{align}\label{eq:mix_min_Dself_CE}
		\frac{1}{D_{{\rm CE},i}^\circ} &= \frac{x_i}{D_{{\rm CE},i}^{{\rm pure},\circ}} + \frac{x_j}{D_{{\rm CE},i}^{\infty,\circ}}\quad {\rm and}\\
		\frac{1}{D_{{\rm CE},j}^\circ} &= \frac{x_j}{D_{{\rm CE},j}^{{\rm pure},\circ}} + \frac{x_i}{D_{{\rm CE},j}^{\infty,\circ}},
	\end{align}
	where the Rosenfeld-scaled self-diffusion coefficients of the pure components $D_{{\rm CE},i}^{{\rm pure},\circ}$ and $D_{{\rm CE},j}^{{\rm pure},\circ}$ are adapted from Ref. \cite{schmitt_entropy_2024} as
	\begin{equation}\label{eq:CE_D_pure}
		D_{{\rm CE},i}^{{\rm pure},\circ} =	\frac{3}{8\sqrt{\uppi}} \frac{1}{\sigma_{i}^2 \Omega_i^{(1,1)}}	\left(T \left(\frac{{\rm d}B}{{\rm d}T}\right) + B \right)^{2/3},
	\end{equation}
	and the Rosenfeld-scaled infinite-dilution diffusion coefficients $D_{{\rm CE},i}^{\infty,\circ}$ and $D_{{\rm CE},j}^{\infty,\circ}$ are given by Eq. (\ref{eq:D_CE_plus}).
	For the Maxwell-Stefan diffusion coefficient in the mixture, the corresponding Chapman-Enskog diffusion coefficient is given as 
	\begin{equation}\label{eq:D_MS_CE_plus}
		\DMS_{{\rm CE},ij}^{\circ} = \frac{3}{8 \sqrt{\uppi}} \frac{1}{\sigma_{ij}^2 \Omega^{(1,1)}_{ij}}	\left(T \left(\frac{{\rm d}B}{{\rm d}T}\right) + B \right)^{2/3},
	\end{equation}
	where $B$ is the second virial coefficient of the mixture calculated as
	\begin{equation}
		B = x_i^2 B_i + x_i x_j B_{ij} + x_j^2 B_j,
	\end{equation}
	where $B_i$ and $B_j$ are the second virial coefficients of the pure components and $B_{ij}$ is cross second virial coefficient.
	
	For determining the minimum of $\Lambda_{\rm CE}^{\circ}$, i.e. $\min \left( \Lambda_{\rm CE}^{\circ} \right)$, the mixing rule proposed by Miller and Carman \cite{miller_self-diffusion_1961} is employed, for both the self-diffusion coefficients as well as for the Maxwell-Stefan diffusion coefficient as
	\begin{align}
		\frac{1}{\min D_{{\rm CE},i}^\circ} &= \frac{x_i}{\min D_{{\rm CE},i}^{{\rm pure},\circ}} + \frac{x_j}{\min D_{{\rm CE},i}^{\infty,\circ}},\\
		\frac{1}{\min D_{{\rm CE},j}^\circ} &= \frac{x_i}{\min D_{{\rm CE},j}^{\infty,\circ}} + \frac{x_j}{\min D_{{\rm CE},j}^{{\rm pure},\circ}},\\
		\frac{1}{\min \DMS_{{\rm CE},ij}^\circ} &= \frac{x_i}{\min D_{{\rm CE},j}^{\infty,\circ}} + \frac{x_j}{\min D_{{\rm CE},i}^{\infty,\circ}}. \label{eq:D_MS_min}
	\end{align}
	
	\section{Simulation details}
	
	Molecular dynamics (MD) simulations were carried out in this work for both model systems as well as real substance systems.
	In both cases, the infinite-dilution diffusion coefficients were sampled. 
	All simulations were carried out with the simulation engine ms2.
	The Lennard-Jones potential between two particles $i$ and $j$ is defined as 
	\begin{equation}
		u_{ij} = 4\varepsilon_{ij} \left[ \left(\frac{\sigma_{ij}}{r_{ij}}\right)^{12} - \left(\frac{\sigma_{ij}}{r_{ij}}\right)^{6} \right],
	\end{equation}
	where $u_{ij}$ is their potential energy, $r_{ij}$ the distance between both particles, and $\sigma_{ij}$ and $\varepsilon_{ij}$ the size and energy parameters of the particles, respectively.
	For all three systems, the size parameters of both components were equal, i.e. $\sigma_2 = \sigma_1$.
	The energy parameter was $\varepsilon_2 = 0.9 \varepsilon_1$ for the first system and $\varepsilon_2 = 0.5 \varepsilon_1$ for the two other systems.
	The potential parameters for the interaction of unlike particles were calculated according to Eqs. (\ref{eq:lorentz}) and (\ref{eq:berthelot}) with $\xi_{12} = 1.2$ for the first system, $\xi_{12} = 1$ for the second system, and $\xi_{12} = 0.85$ for the third system.
	All three systems show a strongly non-ideal behavior, a high-boiling azeotrope (first mixture), a supercritical low-boiling component (second mixture), and a miscibility gap (third mixture).
	
	For each system, simulations at 114 state points in the gas, liquid, metastable vapor-liquid, metastable solid-liquid, and supercritical region (of the solvent) were carried out (see Fig. 2 in the main text).
	For each temperature-pressure pair, three simulations in the vicinity of the infinite dilution limit ($x_2 = 0.001,0.005,0.01\,{\rm mol\,mol^{-1}}$) were carried out. 
	The three simulations were used to extrapolate to infinite dilution ($x_2 \rightarrow \infty$) in a post-processing (see below).
	
\begin{figure}[h]
		\centering
		\includegraphics[width=12cm]{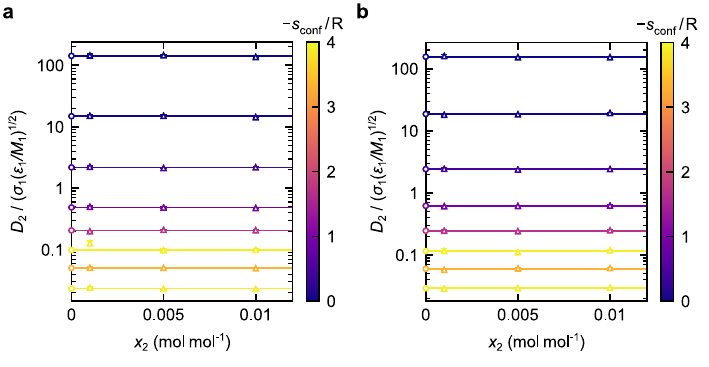}
		\caption{\textbf{Extrapolation to infinite dilution.} Self-diffusion coefficient $D_2$ as function of the mole fraction $x_2$ for eight state points (exemplarily chosen) in two Lennard-Jones systems with a) $\sigma_2 = \sigma_1$, $\varepsilon_2 = 0.9\varepsilon_1$, $\varepsilon_{12} = 1.2\sqrt{\varepsilon_1\varepsilon_2}$ (a) and $\sigma_2 = \sigma_1$, $\varepsilon_2 = 0.5\varepsilon_1$, and $\varepsilon_{12} = 0.85\sqrt{\varepsilon_1\varepsilon_2}$ (b). Triangles are the simulation results for the self-diffusion coefficient $D_2$, the circles are the extrapolated infinite-dilution diffusion coefficients $D_{2}^{\infty}$, and the lines represent the linear extrapolations. The color indicates the configurational entropy. Source data are provided as a Source Data file.}
		\label{fig:extrapolation}
	\end{figure}

	The simulations were performed with the software \textit{ms}2\cite{fingerhut_ms2_2021}.
	Each simulation consisted of 5,000 particles. 
	The Gear-predictor-corrector algorithm was used for time integration with a time step of $\Delta\tau = 0.001 \, \sigma\sqrt{\varepsilon^{-1} M}$.
	The simulations were conducted in the isochoric-isothermal (NVT) ensemble with $10^5$ equilibration time steps and $5\cdot10^6$ production time steps.
	Periodic boundary conditions were applied in all directions.
	The self-diffusion coefficient of component 2 $D_2$ was sampled using the Green-Kubo formalism with a correlation length of $10^4$ time steps for $\rho \geq 0.1\, \sigma_1^{-3}$ and $10^5$ time steps for $\rho < 0.1 \, \sigma_1^{-3}$.
	The infinite-dilution diffusion coefficient $D_{2}^{\infty}$ at a given temperature-density pair was computed by linear extrapolation from the results at finite dilution, cf. Fig. \ref{fig:extrapolation}. 
	The configurational entropy $s_{\rm conf}$ was determined in the simulations using the relation 
	\begin{equation}\label{eq:entropy}
		s_{\rm conf} = \frac{u_{\rm conf}}{T}-\frac{p}{\rho T} - \sum_{i=1}^{2} \frac{\mu_{{\rm conf},i}}{T},
	\end{equation}
	where $u_{\rm conf}$ is the configurational internal energy and $\mu_{{\rm conf},i}$ is the chemical potential of component $i$.
	The chemical potentials were sampled using Widom's test particle method \cite{widom_topics_1963}.
	The results for the system with $\varepsilon_2 = 0.9 \varepsilon_1$ and $\xi_{12} = 0.85$ are presented in the main body of this work. 
	The numeric values for all three studied systems are provided in the electronic Supporting Information. 
	
	For the real substance systems, three mixtures were studied, namely acetone + isobutane, ethanol + chlorine, and benzene + isobutane -- using the same methodology.
	The component-specific force fields from Refs. \cite{guevara-carrion_mutual_2016,eckl_set_2008,stobener_parametrization_2016,schnabel_henrys_2005, windmann_fluid_2014} -- taken from the MolMod database \cite{stephan_molmod_2019} -- were used.
	In total, 29 state points were investigated in the liquid, gas, and supercritical regions for each real substance binary system.
	The time step was $0.329\,{\rm fs}$ for simulations with $\rho > 8.9\,{\rm mol\,L^{-1}}$ and $0.987\,{\rm fs}$ elsewise.
	Each simulation consisted of 4096 molecules -- 4 molecules of the highly diluted component and 4092 molecules of the solvent ($x_{\rm 1} \approx 0.001\,{\rm mol\,mol^{-1}}$).
	The simulations were equilibrated for $5\cdot10^5$ time steps and the actual production run consisted of $6\cdot10^6$ time steps.
	The self-diffusion coefficient of the solvent $D_{1}$ was sampled using the Green-Kubo formalism with a correlation length $10^4$ time steps.
	The chemical potential was sampled using Widom's test particle method to calculate the configurational entropy (see Eq. (\ref{eq:entropy})).
	The second virial coefficient as well as its temperature derivative was sampled for all considered temperatures.
	
	\FloatBarrier
	
	\section{Component-specific EOS models}
	
	The entropy scaling framework proposed in this work for the prediction of mixture diffusion coefficients can be coupled with practically any molecular-based equation of state (and with minor adjustments also with empirical multiparameter EOS \cite{span_multiparameter_2000}, e.g. regarding the molecular property parameters $m$, $\varepsilon$, $\sigma$). 
	In this work, we used the Kolafa-Nezbeda EOS for modeling the Lennard-Jones model mixtures and the PC-SAFT EOS for modeling the real substance mixtures.
	The Kolafa-Nezbeda EOS was found to be the most accurate and robust EOS for modeling thermodynamic properties of the Lennard-Jones fluid \cite{stephan_review_2020,antolovic_phase_2023}. 
	The Kolafa-Nezbeda EOS, originally published for the pure component LJ fluid, was extended in our implementation to mixtures using van der Waals one fluid mixing rule and the modified Lorentz-Berthelot combination rules (cf. Eqs. \ref{eq:lorentz} and \ref{eq:berthelot}). Using the latter, the binary interaction parameter xsi was directly adopted in the EOS model, which makes the theory fully predictive in this case. 
	Also the PC-SAFT EOS is known to often yield good predictions for thermodynamic mixture properties \cite{stephan_vapor-liquid_2024,staubach_modeling_2023}. 
	
	The component-specific PC-SAFT EOS models for the real substances were taken from Refs. \citenum{gross_perturbed-chain_2001,kouskoumvekaki_application_2004,spuhl_reactive_2004,grenner_evaluation_2008,al-saifi_prediction_2008}.
	The parameters are given in Table \ref{tab:eos_parameters}.
	The binary interaction parameters $\xi_{12}$ of all real substance systems was unity for the systems $n$-hexane + $n$-dodecane, toluene + $n$-hexane, and 2-propanol + $n$-heptane; 
	for acetone + chloroform, it was adjusted to $\xi_{12} = 1.02$; 
	for nitrobenzene + $n$-hexane, it was adjusted to $\xi_{12} = 1.05$;
	for acetone + acetonitrile, it was adjusted to $\xi_{12} = 0.97$.
	For the parametrization, only phase equilibrium data was used, i.e. no transport property data. 
	Hence, the mixture diffusion coefficients are described in a predictive way. 
	For applications, the binary interaction parameter could alos be adjusted to experimental diffusion coefficient data to improve the model performance. 
	\begin{table}[h]
		\centering
		\caption{Component-specific PC-SAFT EOS parameters from the literature used in the present work. The columns indicate (from left to right): The substance name, segment diameter $\sigma$, segment dispersion energy $\varepsilon$, chain length parameter $m$, dipole moment $D$, association volume $\kappa_{\rm AB}$, and association strength $\varepsilon_{\rm AB}$.}
		\begin{tabular}{lcccllll}
			\hline
			substance    & $\sigma$ & $\varepsilon/\KB$ &   $m$   & $\mu$ & $\kappa_{\rm AB}$ & $\varepsilon_{\rm AB}/\KB$ & Ref                               \\
			&   \AA    &            K            &         & D     &                   & K                  &                                   \\ \hline
			$n$-hexane   &  3.7983  &         236.77          & 3.0576  &       &                   &                    & \citenum{gross_perturbed-chain_2001} \\
			$n$-heptane  &  3.8049  &         238.40          & 3.4831  &       &                   &                    & \citenum{gross_perturbed-chain_2001} \\
			$n$-dodecane &  3.8959  &         249.21          &  5.306  &       &                   &                    & \citenum{gross_perturbed-chain_2001} \\
			acetone      &  3.2557  &         253.406         & 2.77409 &       &                   &                    & \citenum{kouskoumvekaki_application_2004}                    \\
			acetonitrile &  3.3587  &         313.04          & 2.2661  &       &                   &                    & \citenum{spuhl_reactive_2004}                      \\
			chloroform   &  3.4709  &         271.63          & 2.5038  &       &                   &                    & \citenum{kouskoumvekaki_application_2004}                    \\
			nitrobenzene &  3.6415  &         344.88          & 3.1442  &       &                   &                    & \citenum{grenner_evaluation_2008}    \\
			toluene      &  3.7169  &         285.69          & 2.8149  &       &                   &                    & \citenum{gross_perturbed-chain_2001} \\
			2-propanol   &   3.38   &         212.32          &  2.685  & 1.7   & 0.024675          & 2253.9             & \citenum{al-saifi_prediction_2008}   \\ \hline
		\end{tabular}
		\label{tab:eos_parameters}
	\end{table}
	
	\section{Entropy scaling models}
	
	The entropy scaling limiting case model parameters (pure component and pseudo-pure component) used in this work are reported in Table \ref{tab:es_parameters}.
	Additionally, the references of the experimental data used for the parameter adjustment are given.
	In all cases,  two component-specific parameters were used due to small number of experimental data available (in two cases, one parameter was used).
	In cases where more reference data are available, functions with more parameters can be used, cf. Ref. \cite{schmitt_entropy_2024}.
	\begin{table}[h]
		\centering
		\caption{Component-specific entropy scaling parameters used in the present work. The columns indicate (from left to right): the system, the property, the parameters $\alpha_{2,i}$ and $\alpha_{3,i}$ as well as the reference where experimental data were taken from.}
		\begin{tabular}{lcccc}
			\hline
			system                            &       property       & $\alpha_{2,i}$ & $\alpha_{3,i}$ &                  Ref. data                  \\ \hline
			$n$-hexane (1) + $n$-dodecane (2) &     $D_{\rm 1}$      &    -2.5414     &    -1.9186     & \citenum{suarez-iglesias_self-diffusion_2015} \\
			&     $D_{\rm 2}$      &    -4.3257     &    -3.2885     &                                               \\
			& $D_{\rm 1}^{\infty}$ &      0.0       &    -5.3751     &        \citenum{shieh_transport_1969}         \\
			& $D_{\rm 2}^{\infty}$ &      0.0       &    -3.1396     &                                               \\
			acetone (1) + chloroform (2)      &     $D_{\rm 1}$      &    -0.8935     &    -1.9973     & \citenum{suarez-iglesias_self-diffusion_2015} \\
			&     $D_{\rm 2}$      &    -1.1496     &    -2.0841     &                                               \\
			& $D_{\rm 1}^{\infty}$ &    -3.3309     &    -1.7752     &        \citenum{anderson_mutual_1958}         \\
			& $D_{\rm 2}^{\infty}$ &    -1.9979     &    -1.6190     &                                               \\
			nitrobenzene (1) + $n$-hexane (2) &     $D_{\rm 1}$      &    21.2108     &    -9.9493     & \citenum{suarez-iglesias_self-diffusion_2015} \\
			& $D_{\rm 1}^{\infty}$ &    12.7201     &    -9.3538     &      \citenum{dagostino_prediction_2011}      \\
			toluene (1) + $n$-hexane (2)      & $D_{\rm 1}^{\infty}$ &    -4.0978     &    -1.0285     &       \citenum{chen_determination_1992}       \\
			& $D_{\rm 2}^{\infty}$ &    -2.6050     &    -1.5678     &                                               \\
			toluene (1) + acetonitrile (2)    & $D_{\rm 1}^{\infty}$ &    -0.1904     &    -1.5008     &         \citenum{awan_transport_2001}         \\
			& $D_{\rm 2}^{\infty}$ &    -2.8591     &    -1.7467     &                                               \\
			2-propanol (1) + $n$-heptane (2)  & $D_{\rm 1}^{\infty}$ &    -11.9703    &     0.2606     &    \citenum{andanson_quantification_2016}     \\
			& $D_{\rm 2}^{\infty}$ &     5.4598     &    -2.9687     &                                               \\ \hline
		\end{tabular}
		\label{tab:es_parameters}
	\end{table}
	
	\FloatBarrier
	
	\section{Comparison of the scaling behavior of pure component self-diffusion coefficients and infinite-dilution diffusion coefficients}
	
	For the Lennard-Jones model systems, a quasi-universality is observed using the applied scaling, which is evident by the representation of $\widehat{D}_{2}^{\infty,\circ}$ by the global model from Ref.~\citenum{schmitt_entropy_2024}, which was only fitted to pure-component self-diffusion coefficient data. 
	Hence, the correlation developed in Ref.~\citenum{schmitt_entropy_2024} using pure-component self-diffusion coefficient data of the Lennard-Jones fluid also describes the infinite-dilution diffusion coefficient data from this work.
	
\begin{figure}[h]
		\centering
		\includegraphics[width=12cm]{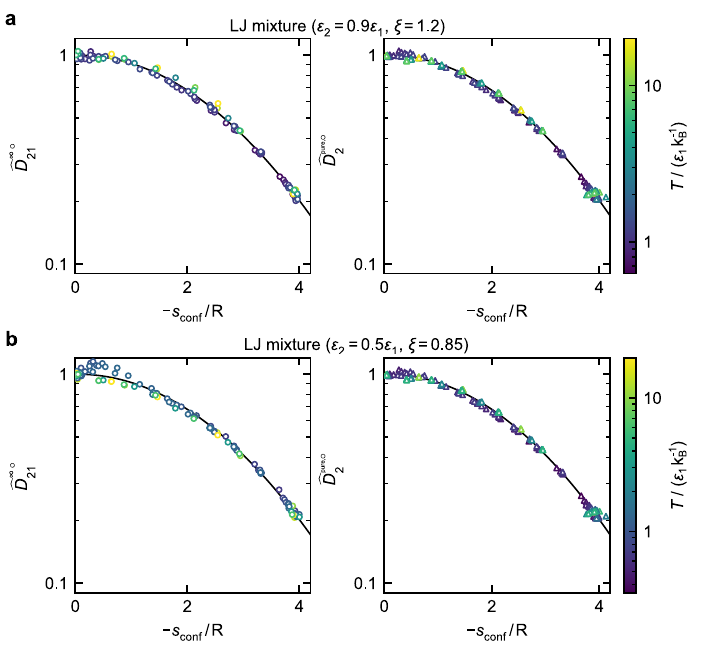}
		\caption{\textbf{Entropy scaling of infinite-dilution diffusion coefficients.} Scaled infinite-dilution diffusion coefficient $\widehat{D}_{2}^{\infty,\circ}$ (left) and self-diffusion coefficient $\widehat{D}_{2}^{{\rm pure},\circ}$ (right) for two Lennard-Jones systems (a and b) as a function of the configurational entropy $s_{\rm conf}$. The symbols are simulation results from this work ($\widehat{D}_{2}^{\infty,\circ}$) and from Ref. \citenum{schmitt_entropy_2024} ($\widehat{D}_{2}^{{\rm pure},\circ}$). The color of the symbols indicates the temperature. The black solid line is the entropy scaling model. Source data are provided as a Source Data file.}
		\label{fig:scaling_comparison}
	\end{figure}

	In Fig. \ref{fig:scaling_comparison}, the scaling of the pure component self-diffusion coefficient is compared with the scaling of the infinite-dilution diffusion coefficient -- for two of the studied LJ systems.
	Both CE-scaled diffusion coefficients show a monovariate behavior over the entire range of configurational entropies and all temperatures.
	To quantify this, a measure $\Delta_{\Lambda}$ was defined as
	\begin{equation}
		\Delta_\Lambda = \sqrt{ \frac{1}{N_{\Lambda}} \sum_{i}^{\rm N_{\Lambda}} (\Lambda_{{\rm sim},i} - \Lambda_{{\rm mod},i})^2 },
	\end{equation}
	where $\Lambda \in \{ \widehat{D}_2^{{\rm pure},\circ}, \widehat{D}_{2}^{\infty,\circ} \}$, $N_{\Lambda}$ is the respective number of simulation data points, $\Lambda_{{\rm sim},i}$ are diffusion coefficients obtained from the simulations, and $\Lambda_{{\rm mod},i}$ the values calculated from the entropy scaling model.
	The obtained values are $\Delta_{\widehat{D}_2^{{\rm pure},\circ}} \approxeq 1.244$ and $\Delta_{\widehat{D}_{2}^{\infty,\circ}} \approxeq 1.254$ for first system and  $\Delta_{\widehat{D}_2^{{\rm pure},\circ}} \approxeq 1.244$ and $\Delta_{\widehat{D}_{2}^{\infty,\circ}} \approxeq 1.254$ for the second Lennard-Jones system.
	Hence, the infinite-dilution diffusion coefficient data yields a monovariate behavior -- with the same extent as the pure component self-diffusion coefficient. 
	This support the perspective that the infinite dilution state can be considered as a pseudo-pure component. 
	
	\section{Scaling behavior of the infinite-dilution diffusion coefficient}
	
\begin{figure}[h!]
		\centering
		\includegraphics[width=12cm]{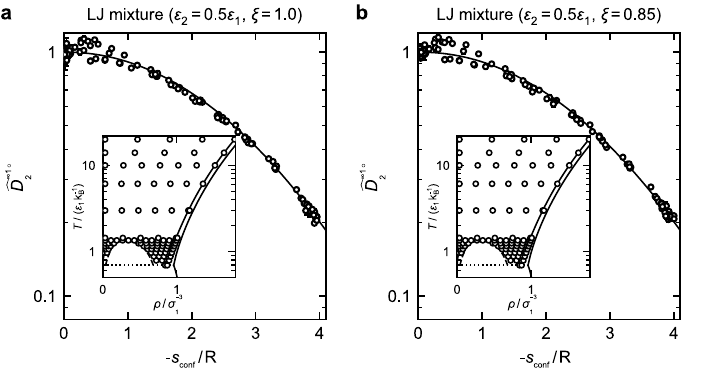}
		\caption{\textbf{Entropy scaling of infinite-dilution diffusion coefficients in two Lennard-Jones mixtures.} a) $\varepsilon_2 = 0.5\,\varepsilon_1$, $\varepsilon_{12} = 1\,\xi_{12}$; b) $\varepsilon_2 = 0.5\,\varepsilon_1$, and $\varepsilon_{12} = 0.85\,\xi_{12}$ (both: $\sigma_2 = \sigma_1$). Scaled infinite-dilution diffusion coefficient of component 2 $\widehat{D}^{\infty,\circ}_{2}$ as a function the reduced configurational entropy $s_{\rm conf}/{\rm R}$. The line indicates the entropy scaling model. Symbols are MD simulation data from this work. The inset shows the simulation state points in the temperature-density phase diagram of component 1. Therein, solid lines indicate the phase envelopes from Refs. \citenum{stephan_thermophysical_2019,schultz_comprehensive_2018}. Source data are provided as a Source Data file.}
		\label{fig:infdilution_lj_DF}
	\end{figure}

	Fig. \ref{fig:infdilution_lj_DF} shows the simulation results for two additional Lennard-Jones mixture.
	The results confirm the findings of the main part: in both mixtures, the scaled infinite-dilution diffusion coefficient shows a monovariate function with respect to the scaled configurational entropy.
	
\begin{figure}[h]
		\centering
		\includegraphics[width=6cm]{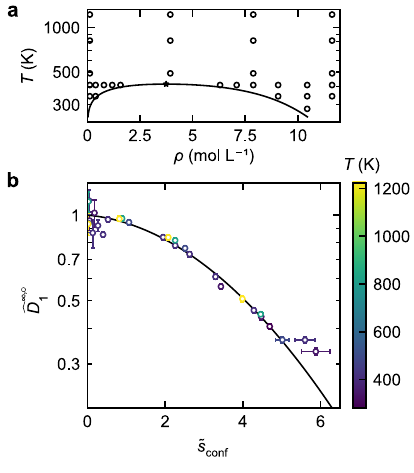}
		\caption{\textbf{Scaling behavior of the infinite-dilution diffusion coefficient of benzene in the system benzene (1) + isobutane (2)}. a) Simulation state points (symbols) in the temperature-density phase diagram of the solvent isobutane. The line indicates the vapor-liquid equilibrium and the star the critical point as calculated from the PC-SAFT EOS \cite{gross_perturbed-chain_2001}. b) Scaled diffusion coefficient of benzene infinitely diluted in isobutane $\widehat{D}_{\rm 1}^{\infty,\circ}$ as a function of the configurational entropy $\tilde{s}_{\rm conf}$. The symbols are the simulation results (color indicates the temperature) and the line the entropy scaling model (fitted to the simulation results). Source data are provided as a Source Data file.}
		\label{fig:infdilution_real}
	\end{figure}

	Fig. \ref{fig:infdilution_real} shows the simulation results for the real substance system benzene + isobutane.
	The simulation state points cover the gas, liquid, and the supercritical region.
	The scaled infinite-dilution diffusion coefficient of benzene $\widehat{D}_{\rm 1}^{\infty,\circ}$ shows a monovariate behavior with respect to the reduced configurational entropy $\tilde{s}_{\rm conf}$.
	The results confirm the validity of the entropy scaling methodology introduced in this work for real substance systems.
	
\begin{figure}[h]
		\centering
		\includegraphics[width=10cm]{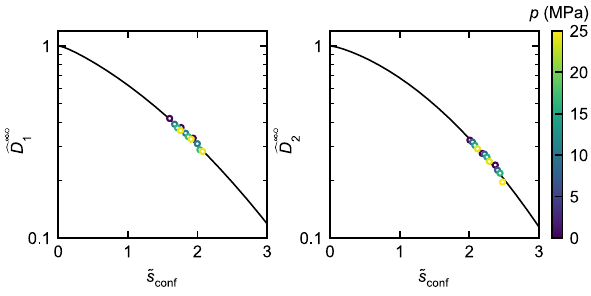}
		\caption{\textbf{Scaling behavior of the infinite-dilution diffusion coefficients in the system toluene (1) + \textit{n}-hexane (2) as function of the reduced configurational entropy $\tilde{s}_{\rm conf}$.} Left: Scaled diffusion coefficient of toluene infinitely diluted in $n$-hexane $\widehat{D}_1^{\infty \circ}$. Right: Scaled diffusion coefficient of $n$-hexane infinitely diluted in toluene $\widehat{D}_2^{\infty \circ}$. The line indicates the system-specific entropy scaling model (parameters given in Table \ref{tab:es_parameters}). The color indicates the pressure. Source data are provided as a Source Data file.}
		\label{fig:infdilution_real_hex-tol}
	\end{figure}

	Experimental infinite dilution self-diffusion coefficient data in a large range of states (as considered for the model systems) are unfortunately not available. 
	For the system toluene + $n$-hexane, experimental data of infinite-dilution diffusion coefficients at different pressures and temperatures are available, but only for the liquid phase. 
	Fig. \ref{fig:infdilution_real_hex-tol} shows the scaled infinite-dilution diffusion coefficients of the system toluene (1) + $n$-hexane (2) as function of the reduced configurational entropy, i.e. $\widehat{D}_1^{\infty, \circ}(\tilde{s}_{\rm conf})$ and $\widehat{D}_2^{\infty, \circ}(\tilde{s}_{\rm conf})$.
	Additionally, the entropy scaling models (lines) are shown for both cases.
	For each case, two system-specific parameters were adjusted (see Table \ref{tab:es_parameters}).
	For the system toluene + $n$-hexane, experimental data for both infinite-dilution diffusion coefficients are available at different temperatures and pressures (which is very seldom).
	Thus, this system is well suited to demonstrate the scaling of the infinite-dilution diffusion coefficients.
	All scaled diffusion coefficients lie on a single curve, i.e. show a monovariate behavior with respect to $\tilde{s}_{\rm conf}$.
	
	\FloatBarrier
	
	\section{Additional entropy scaling results of the Lennard-Jones mixtures}
	
\begin{figure}[h]
		\centering
		\includegraphics[width=6cm]{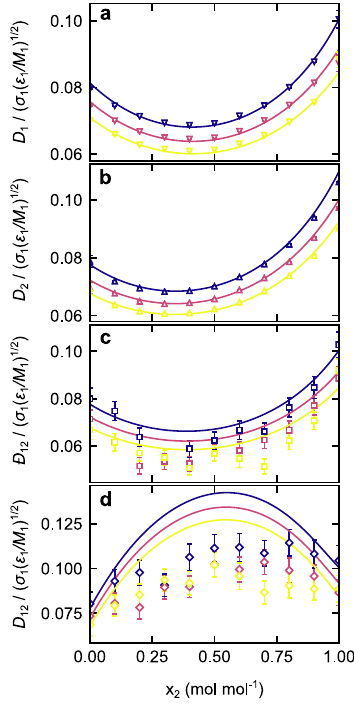}
		\caption{\textbf{Entropy scaling predictions at different pressures.} Diffusion coefficients in the Lennard-Jones system with $\sigma_2 = \sigma_1$, $\varepsilon_2 = 0.9\varepsilon_1$, and $\varepsilon_{12} = 1.2\sqrt{\varepsilon_1\varepsilon_2}$ as a function of the mole fraction $x_2$ at $T = 0.92\,\KB\varepsilon_1^{-1}$. a) Self-diffusion coefficient of component 1 $D_1$; b) Self-diffusion coefficient of component 2 $D_2$; c) Maxwell-Stefan diffusion coefficient $\DMS_{12}$; d) Fickian diffusion coefficient $D_{12}$. Lines are the predictions from the entropy scaling model. Symbols are simulation results from Ref. \citenum{fertig_influence_2023}. The entropy scaling model was used in combination with the Kolafa-Nezbeda EOS\cite{kolafa_lennard-jones_1994}. The colors indicate the pressure $p \in \{ 0.13,0.26,0.39 \}\,\sigma^3_1\varepsilon_1^{-1}$ (yellow to dark purple). Source data are provided as a Source Data file.}
		\label{fig:mix_B_T_iso}
	\end{figure}

	Fig. \ref{fig:mix_B_T_iso} shows the application of the entropy scaling model to the Lennard-Jones mixture with $\varepsilon_2 = 0.9\varepsilon_1$ and $\varepsilon_{12} = 1.2\sqrt{\varepsilon_1\varepsilon_2}$ at different pressures (corresponding to Fig. 4 of the main body).
	The predictions from the entropy scaling model are compared to simulation data from Ref. \citenum{fertig_influence_2023}.
	The results for different pressures are very similar to those for different temperatures (see main body of this work):
	The agreement between the predictions from the entropy scaling model and simulation data is very good for both self-diffusion coefficients.
	For the mutual diffusion coefficients, some deviations are observed.
	
	\FloatBarrier
	
	\section{Comparison of entropy scaling to the Vignes and Darken models}
	
	The performance of the entropy scaling model is compared to established empirical models. 
	The Vignes equation \cite{vignes_diffusion_1966} is an often applied, simple model for calculating Maxwell-Stefan diffusion coefficients in mixtures based on the infinite-dilution diffusion coefficients.
	It is written as
	\begin{equation}\label{eq:vignes}
		\DMS_{ij} = \left( D_{i}^{\infty} \right)^{x_j} \left( D_{j}^{\infty} \right)^{x_i}.
	\end{equation}
	Besides the Vignes model, the generalized Darken model\cite{krishna_darken_2005} is often applied and found to be superior in some cases \cite{liu_predictive_2011}.
	For binary mixtures, it is defined as
	\begin{equation}\label{eq:darken}
		\DMS_{ij} = x_i \left( x_i^{\rm (m)} D_{j}^{\infty} + x_j^{\rm (m)} D_{j}^{\rm (pure)} \right) + x_j \left( x_i^{\rm (m)} D_i^{\rm (pure)} + x_j^{\rm (m)} D_{i}^{\infty} \right),
	\end{equation}
	where $x_i^{\rm (m)}$ and $x_j^{\rm (m)}$ are the mass fractions of the components $i$ and $j$, respectively.
	Both the Vignes and the Darken model only require information on the limiting case diffusion coefficients, like the entropy scaling model proposed in this work.
	
\begin{figure}[h]
		\centering
		\includegraphics[width=6cm]{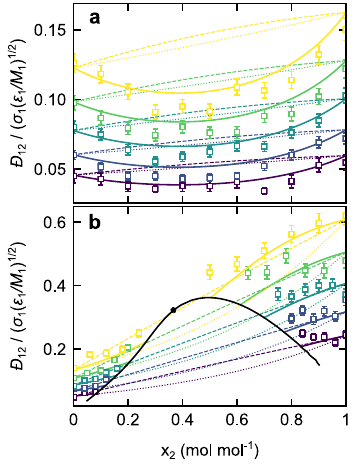}
		\caption{\textbf{Comparison of MD reference data to the entropy scaling model proposed in this work, the Vignes model, and the Darken model for the Maxwell-Stefan diffusion coefficient.} Results for two binary Lennard-Jones systems (a: $\sigma_2 = \sigma_1$, $\varepsilon_2 = 0.9\varepsilon_1$, $\varepsilon_{12} = 1.2\sqrt{\varepsilon_1\varepsilon_2}$, b: $\sigma_2 = \sigma_1$, $\varepsilon_2 = 0.5\varepsilon_1$, $\varepsilon_{12} = 0.85\sqrt{\varepsilon_1\varepsilon_2}$) as a function of the mole fraction $x_2$ at $p = 0.13\,\sigma^3_1\varepsilon_1^{-1}$ (a) and $p = 0.26\,\sigma^3_1\varepsilon_1^{-1}$ (b). Symbols indicate simulation results from Ref. \citenum{fertig_influence_2023}. Solid lines are predictions from the entropy scaling model obtained in combination with the Kolafa-Nezbeda EOS. Dotted lines are results from the Vignes model (cf. Eq. (\ref{eq:vignes})) and dashed lines from the generalized Darken model (cf. Eq. (\ref{eq:darken})). The colors indicate the temperature $T \in \{ 0.79,0.855,0.92,0.985,1.05 \}\,\KB\varepsilon_1^{-1}$ (blue to yellow). The black line indicate the liquid-liquid equilibrium diffusion coefficient and the star the critical point. Error bars represent the simulation uncertainty given in Ref. \cite{fertig_influence_2023}. Source data are provided as a Source Data file.}
		\label{fig:vignes}
	\end{figure}

	In Fig. \ref{fig:vignes}, the predictions from the entropy scaling model proposed in this work are compared to the results from the Vignes model and the generalized Darken model for both Lennard-Jones mixtures.
	Both mixtures show a strongly non-ideal behavior (see main text and above).
	For both mixtures, the entropy scaling model provides a reasonable description of the Maxwell-Stefan diffusion coefficient in the mixture.
	The Vignes and Darken models are not able to capture the trend of $\DMS_{12}$ in the first Lennard-Jones system (see Fig. \ref{fig:vignes}a).
	For the second system (see Fig. \ref{fig:vignes}b), the predictions by the Vignes equation show a wrong curvature compared to the simulation data and, most importantly, do not capture the liquid-liquid equilibrium (LLE). 
	Both empirical models (Vignes and Darken) do not comprise information on the liquid-liquid miscibility gap. 
	The entropy scaling model proposed in this work, on the other hand, inherently captures the LLE due to the coupling with the EOS model and also describes the diffusion coefficients of the coexisting phases, metastable phases, supercritical phases etc.
	However, no computer experiment data is available for the coexisting phase diffusion coefficients for validation. 
	
	\FloatBarrier
	
	\section{Scaled diffusion coefficients in mixtures}
	
\begin{figure}[h]
		\centering
		\includegraphics[width=12cm]{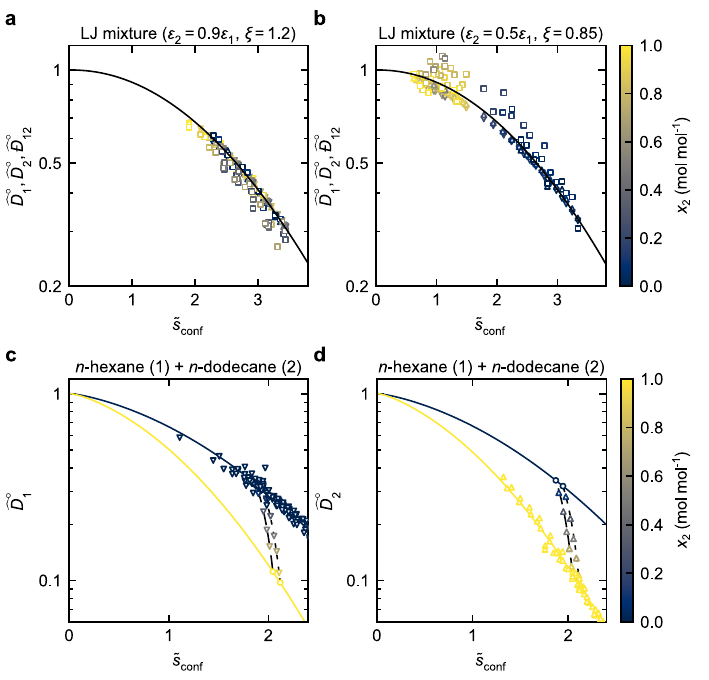}
		\caption{\textbf{Scaled diffusion coefficients as a function of the configurational entropy $\tilde{s}_{\rm conf}$ in different mixtures.} a and b: Lennard-Jones mixtures. c and d: Mixture $n$-hexane (1) + $n$-dodecane (2). Symbols are scaled simulation results (a and b) or experimental data (c and d) \cite{fertig_transport_2022,shieh_transport_1969}. Triangles: Self-diffusion coefficients; squares: Maxwell-Stefan diffusion coefficient; circles: infinite-dilution diffusion coefficients. a + b: The solid line represents the global entropy scaling model for the diffusion coefficients of the Lennard-Jones fluid \cite{schmitt_entropy_2024}. c: The yellow and the dark blue lines represent the entropy scaling models for $D_{1}$ and $D_{1}^{\infty}$, respectively. d: The dark blue and the yellow lines represent the entropy scaling model for $D_{2}$ and $D_{2}^{\infty}$, respectively. c and d: Black dotted and dashed lines are results from the entropy scaling model for constant temperature (dotted: $T=298.15\,K$, dashed: $T=308.15\,{\rm K}$) computed over the entire composition range, i.e. $0 < x_2 \,/\,{\rm mol\,mol^{-1}} < 1$. Source data are provided as a Source Data file.}
		\label{fig:scalings_mix}
	\end{figure}

	Fig. \ref{fig:scalings_mix} shows the scaling behavior of diffusion coefficients in mixtures for the two Lennard-Jones systems and a real substance system.
	For the two considered Lennard-Jones systems, both scaled self-diffusion coefficients and the scaled Maxwell-Stefan diffusion coefficient collapse on one line, which is a special feature of the considered Lennard-Jones mixtures.
	As a result, a single set of parameters is able to describe all three diffusion coefficients.
	The scattering of the Maxwell-Stefan diffusion coefficients is larger than that for the self diffusion coefficients, which is due to the scattering of the reference data.
	For the real substance system $n$-hexane (1) + $n$-dodecane (2), results for both scaled self-diffusion coefficients $\widehat{D}^{\circ}_{1}$ and $\widehat{D}^{\circ}_{2}$ are shown.
	The corresponding, non-scaled results are shown in Fig. 5a of the main text.
	The data and models for the pure component self-diffusion coefficient and the pseudo-pure component infinite-dilution diffusion coefficient differ significantly. 
	The self-diffusion coefficients in the mixture ($0 < x_{\rm 2}\,/ \, {\rm  mol\,mol^{-1}} <1$) lie between both curves.
	The entropy scaling model connects both lines (here at a given $T$ and $p$) and is thus able to predict these points.
	The link is primarily established via the entropy of the mixture and the mixing and combination rules built in the entropy scaling model.
	
	Fig. \ref{fig:scalings_mix} demonstrates the two central elements of the proposed methodology: 
	(1) the infinite-dilution diffusion coefficients, if scaled as proposed in this work, exhibit a monovariate relation and can be treated as a pseudo-pure component, which enables the scaling of that property. 
	The scaling can be used for predicting infinite-dilution diffusion coefficients far beyond the range of available data based on that scaling. 
	(2) The predictions of the diffusion coefficients into the mixture do not follow the monovariate scaling behavior. 
	Yet, the entropy of the mixture in combination with appropriately designed mixing and combination rules enable the prediction of the diffusion coefficients in the mixture in the scaled variables. 
	For the limiting cases of the pure components and the pseudo-pure components, empirical models for describing the scaled diffusion coefficient are required in this framework. 
	Yet, these models require only very few parameters, e.g. 1 or 2 parameters were used for the real substance cases studied in this work (cf. Table \ref{tab:es_parameters}). 
	From these few parameters, the model (1) can predict the corresponding diffusion coefficient practically in the entire fluid state region, cf. for example Fig. \ref{fig:scaling_comparison}. 
	For the main part of the novel framework, i.e. the prediction of the different diffusion coefficients in the mixture $D_{i}(x_j)$, $D_{j}(x_j)$, $\DMS_{ij}(x_j)$, and $D_{ij}(x_j)$ no adjustable parameters are required. 
	The mechanisms for establishing the link between the limiting case diffusion coefficients (pure component and pseudo-pure component) are analogue and consistent to the mechanisms usually used for predicting the viscosity and thermal conductivity of mixtures by entropy scaling. 
	Albeit, significantly more complex in the case of diffusion since different diffusion coefficients are described in a single and consistent framework.
	However, strong non-idealities of the Maxwell-Stefan diffusion coefficients, especially in binary mixtures of an alcohol and a non-polar substance \cite{rutten_diffusion_1992}, may not be covered by the proposed framework.
	
\begin{figure}[h!]
		\centering
		\includegraphics[width=6cm]{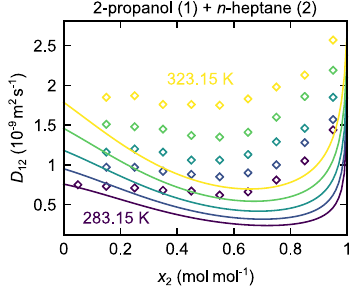}
		\caption{\textbf{Fickian diffusion coefficients of the mixture 2-propanol (1) + \textit{n}-heptane (2).} Predictions by the entropy scaling model as a function of the mole fraction $x_2$ at $p = 0.1\,{\rm MPa}$. Symbols are experimental data from Ref. \citenum{he_mutual_2016} and lines are model predictions. Source data are provided as a Source Data file.}
		\label{fig:real_mutual_alc}
	\end{figure}

	\reffig{fig:real_mutual_alc} shows results for a system with an associating component, namely 2-propanol + $n$-heptane. 
	For associating components, entropy scaling is known to often perform less good \cite{lotgering-lin_pure_2018,schmitt_entropy_2024}.
	This is also observed here.
	For the system 2-propanol + $n$-heptane, significant quantitative deviations between entropy scaling model and the experimental data are observed ($\overline{\delta D_{12}} \approx 50\,\%$).
	Nevertheless, qualitatively, the minimum of $D_{12}(x_{2})$ and the temperature dependency are correctly captured by the model. 
	The entropy scaling shows a prediction with a distinct minimum at $x_2 \approx 0.7 \; {\rm mol \, mol^{-1}}$ while the experimental data only show slightly non-ideal behavior.
	The deviations are possibly due to several reasons. 
	The mutual diffusion coefficients of system consisting of an associating and a non-associating component obey a complex behavior at low concentrations of the associating component (here: 2-propanol) \cite{rutten_diffusion_1992} which is due to a structuring in the liquid.
	Additionally, the modeling of other properties in such systems by equations of state also posses challenges. 
	The reasons for the deviations observed in Fig~\ref{fig:real_mutual_alc} are manifold and require further investigations.
	The description of these non-idealities, which might be due to local composition phenomena \cite{li_mutual-diffusion-coefficient_2001}, requires modifications of the introduced framework as well as of the underlying EOS models.
	
	\FloatBarrier
	
	
	\section{Diffusion coefficients at metastable and unstable states}
	
\begin{figure}[h]
		\centering
		\includegraphics[width=12cm]{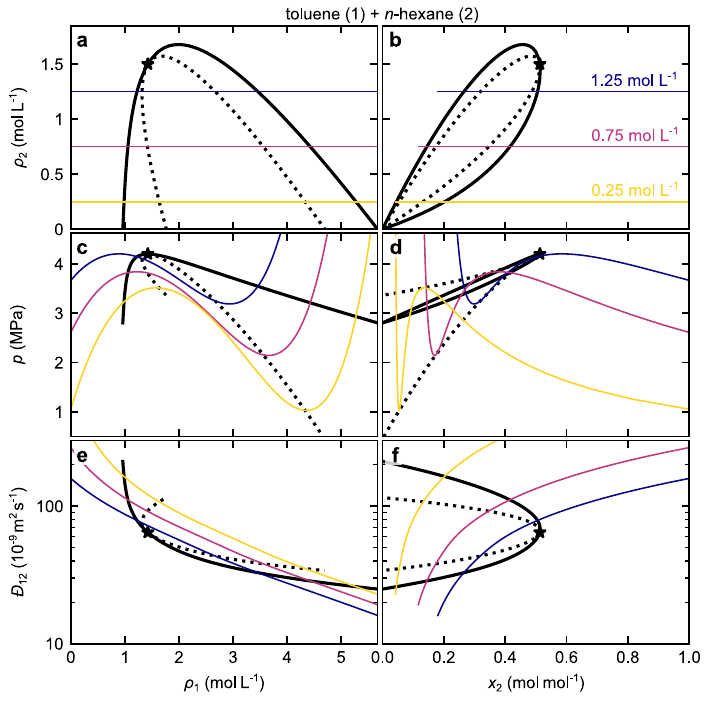}
		\caption{\textbf{Predictions of the VLE properties and the Maxwell-Stefan diffusion coefficient from the entropy scaling + EOS model in the two-phase vapor-liquid equilibrium region of the binary mixture toluene (1) + \textit{n}-hexane (2).}
			Partial density of $n$-hexane $\rho_{2}$ (top), pressure $p$ (middle), and Maxwell-Stefan diffusion coefficient $\DMS_{12}$ (bottom) as a function of the partial density of toluene $\rho_{1}$ (a, c, and e) and of the mole fraction $x_{2}$ (b, d, and f) at $T = 560\,{\rm K}$. The black solid lines correspond to the vapor-liquid binodal, the dotted line to the vapor-liquid spinodal, and the black star indicates the critical point. The three colored lines are lines with constant partial density of $n$-hexane $\rho_{2}$. Source data are provided as a Source Data file.}
		\label{fig:mix_metastable}
	\end{figure}

	Fig. \ref{fig:mix_metastable} demonstrates the application of the entropy scaling model to diffusion coefficients at metastable and unstable states in mixtures.
	Therefore, three lines at constant partial densities of $n$-hexane $\rho_{2} = 0.25,\;0.75,\;{\rm or}\;1.25\,{\rm mol\,L^{-1}}$ were calculated by varying the partial density of toluene in the range $0\,{\rm mol\,L^{-1}} \leq \rho_{1} \leq 5.6\,{\rm mol\,L^{-1}}$ at a temperature $T = 560\,{\rm K}$, where $n$-hexane is supercritical.
	This procedure is computationally convenient for calculating metastable and unstable states as the applied PC-SAFT EOS is formulated in the Helmholtz energy $a$ with its fundamental variables $T$, $\rho$, and $x$.
	All three partial isochores cross the vapor-liquid equilibrium.
	The corresponding pressures undergo a van der Waals loop with minima and maxima at the spinodals (see Fig. \ref{fig:mix_metastable}).
	Evidently, the application of the entropy scaling model in metastable states requires an EOS model that shows a physically reasonable behavior in that region, i.e. a single van der Waals loop, which is the case here. 
	Here, only predictions for the Maxwell-Stefan diffusion coefficient are shown.
	The Maxwell-Stefan diffusion coefficient smoothly transitions from the liquid phase to the gas phase through the vapor-liquid equilibrium including the metastable and unstable regions.
	While there is no diffusion coefficient reference data available for validation, these predictions seem physically reasonable. 
	
	\newpage
	\FloatBarrier
	\renewcommand{\refname}{Supplementary References}